\documentclass[12pt,a4paper,oneside]{article}
\usepackage{graphicx,amssymb,amsmath,color}
\usepackage[linkcolor={black},citecolor={red},colorlinks=true]{hyperref}
\usepackage{cite}
\usepackage{enumerate}
\usepackage[margin=2 cm]{geometry}
\usepackage{slashed}
\usepackage{multirow}
\usepackage[normalem]{ulem}
\usepackage[dvipsnames, x11names]{xcolor}
\usepackage[compat=1.0.0]{tikz-feynman}
\usepackage[labelfont=bf]{caption}
\usepackage{subfig}

\usepackage{cleveref}
\usepackage{cancel}

\definecolor{flag}{rgb}{1.0, 0.03, 0.0}

\def\mt{\tilde{m}}

\def\kt{\tilde{k}}
\def\vt{\tilde{v}}
\def\kp{k_\perp}

\def\beq{\begin{equation}}
\def\eeq{\end{equation}}
\def\bea{\begin{eqnarray}}
\def\eea{\end{eqnarray}}

\def\bit{\begin{itemize}}
\def\eit{\end{itemize}}

\def\l{\left}
\def\r{\right}

\def\baa{\begin{array}}
\def\eaa{\end{array}}

\def\d{\partial}

\def\simgt{\mathrel{\lower2.5pt\vbox{\lineskip=0pt\baselineskip=0pt
           \hbox{$>$}\hbox{$\sim$}}}}
\def\simlt{\mathrel{\lower2.5pt\vbox{\lineskip=0pt\baselineskip=0pt
           \hbox{$<$}\hbox{$\sim$}}}}
\newcommand{\vev}[1]{ \langle {#1} \rangle }

\def\bfc{\begin{figure}\begin{center}}
\def\efc{\end{center}\end{figure}}
\def\nn{\nonumber\\}

\definecolor{chromeyellow}{rgb}{1.0, 0.65, 0.0}
\definecolor{darkcoral}{rgb}{0.8, 0.36, 0.27}
\definecolor{cadmiumgreen}{rgb}{0.0, 0.42, 0.24}

\begin{document}

\begin{flushright}
\hspace{3cm} 
SISSA  10/2024/FISI
\end{flushright}
\vspace{.6cm}
\begin{center}

{\Large \bf 
{NLO friction in symmetry restoring phase transitions}}
\vspace{0.5cm}

\vspace{1cm}{Aleksandr Azatov$^{a,b,c,1}$, Giulio Barni$^{a,b,2}$, Rudin Petrossian-Byrne$^{d,3}$ }
\\[7mm]
 {\it \small
$^a$ SISSA International School for Advanced Studies, Via Bonomea 265, 34136, Trieste, Italy\\
$^b$ INFN - Sezione di Trieste, Via Bonomea 265, 34136, Trieste, Italy\\[0.1cm]
$^c$ IFPU, Institute for Fundamental Physics of the Universe, Via Beirut 2, 34014 Trieste, Italy\\[0.1cm]
$^d$ Abdus Salam International Centre for Theoretical Physics, Strada Costiera 11, 34151, Trieste, Italy\\[0.1cm]
}
\end{center}

\bigskip \bigskip \bigskip

\centerline{\bf Abstract} 
  \begin{quote}

Interactions between bubbles/domain walls and the surrounding medium are a topic of active research, particularly as they apply to friction effects on accelerated expansion during first-order phase transitions. In this paper, we analyze for the first time friction pressure on relativistic walls in phase transitions where gauge symmetry is restored, particularly motivated by the observation that this pressure can, in principle, be negative at leading order, since some particles lose mass by definition as they cross into the new phase.
We find, however, that at NLO, the soft emission of vectors from a charged current leads to positive pressure scaling as the wall's Lorentz boost factor $\gamma_w$, similar to the case of gauge symmetry breaking. Contrary to the latter case, we find that the dominant contribution in single emission is safe from IR divergences and exhibits a much stronger dependence on the wall shape. Finally, we argue that in any phase transition, no multi-particle process on the wall can impart negative pressure greater than the leading order result, in the asymptotic limit of large velocity.
\end{quote}

\vfill
\noindent\line(1,0){188}
{\scriptsize{ \\ E-mail:
\texttt{$^1$\href{mailto:aleksandr.azatov@NOSPAMsissa.it}{aleksandr.azatov@sissa.it}}, \texttt{$^2$\href{mailto:giulio.barni@NOSPAMsissa.it}{giulio.barni@sissa.it}, \texttt{$^3$\href{mailto:rpetross@NOSPAMictp.it}{rpetross@ictp.it}}
}}}

\newpage
\tableofcontents
\section{Introduction}
First order phase transitions (FOPT) are violent out-of-equilibrium phenomena that might have occurred in the history of our universe \cite{PhysRevD.15.2929,PhysRevD.16.1762, LINDE1983421}, with a vast variety of phenomenological consequences for baryogenesis \cite{Kuzmin:1985mm, Shaposhnikov:1986jp,Nelson:1991ab,Carena:1996wj,Cline:2017jvp,Long:2017rdo,Bruggisser:2018mrt,Bruggisser:2018mus,Morrissey:2012db,Azatov:2021irb, Baldes:2021vyz,Fernandez-Martinez:2022stj, Chun:2023ezg}, dark matter \cite{Falkowski:2012fb, Baldes:2020kam, Azatov:2021ifm,Baldes:2021aph, Asadi:2021pwo, Baldes:2022oev, Baldes:2023fsp}, black hole production \cite{10.1143/PTP.68.1979,Kawana:2021tde,Jung:2021mku,Gouttenoire:2023naa,Lewicki:2023ioy}, and stochastic 
gravitational wave backgrounds \cite{Witten:1984rs,Hogan_GW_1986,Kosowsky:1992vn,Kosowsky:1992rz,Kamionkowski:1993fg} potentially observable at present and future detectors \cite{Caprini:2015zlo, Caprini:2019egz}. To make quantitative predictions for any of these consequences, it is important to properly understand all stages of the process: bubble nucleation, expansion and percolation. This work relates to the middle stage.

\begin{figure}
\centering
    \includegraphics[width=.5\textwidth]{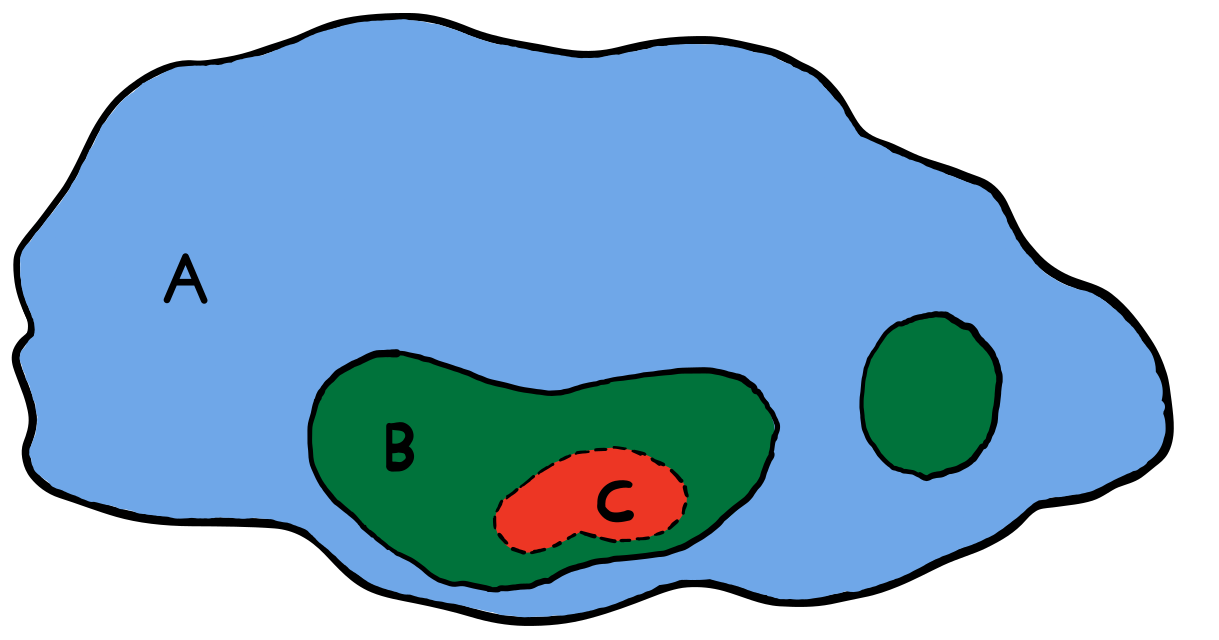}
    \caption{In the vast landscape of quantum field theories A, a subset B can undergo vacuum decay and first order phase transitions. Within B we can distinguish between phase transitions where nucleated bubbles can in principle accelerate forever, and that subset C where instead an equilibrium velocity always exists because friction from surrounding matter grows indefinitely with $\gamma$. The exact boundaries of C within B are a topic of ongoing research. The current work extends C to include gauge symmetry restoring transitions with charged currents. Clearly, we have not attempted to draw the relative sizes of A, B and C to scale.}
    \label{fig:landscapes}
\end{figure}

Approximate bubble nucleation rates are computed from standard equilibrium dynamics \cite{PhysRevD.15.2929,PhysRevD.16.1762, LINDE1983421}. 
Once nucleated, a bubble is a classical solution to the relevant field equations that, if unimpeded,  will expand to relativistic speeds according to
\begin{equation}
    \dot{\gamma_w} \approx \frac{\Delta V - \mathcal{P}(\gamma_w)}{\sigma}\ ,
\end{equation}
where $\gamma_w$ is its Lorentz boost factor, $\Delta V$ the energy density difference between the two phases, $\sigma$ the surface tension of the bubble wall, and $\mathcal{P}$ is the \textit{out of equilibrium} pressure exerted on the expanding wall sourced by interactions with surrounding matter. The latter is in general dependent on the details of the theory. If $\Delta V = \mathcal{P}(\gamma_{\rm eq})$ for some $\gamma_{\rm eq}$, then bubbles can reach a constant equilibrium velocity and most of the energy density released is transferred to the medium \footnote{It is sometimes possible for bubbles to collide for some $\gamma_w < \gamma_{\rm eq}$, in which case the existence, in principle, of the equilibrium value is somewhat meaningless.}. Otherwise, they continue to accelerate and $\Delta V$ goes into the kinetic energy of the walls (`runaway' scenario). 
If $\mathcal{P}(\gamma_w\rightarrow\infty) \rightarrow \infty$ the theory always admits an equilibrium. The space of theories with this feature is a topic of recent and ongoing research, as sketched in \cref{fig:landscapes}.
All phenomenological consequences listed above depend sensitively on the bubble speed.

\begin{figure}
\centering
    \includegraphics[width=.4\textwidth]{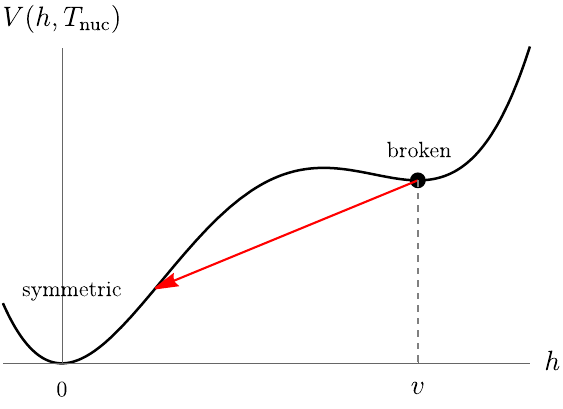}
    \caption{In this paper we focus on friction effects on nucleated bubbles of restored gauge symmetry expanding into a surrounding false vacuum of broken symmetry. The effective potential for the order parameter $h$ is thus of the form sketched, for some $T_{\rm nuc}$.}
    \label{fig:sketch_symm_restoring}
\end{figure}

In this paper, we take a step forward in charting this space by studying the particular case of phase transitions (PT) where gauge symmetry is restored. This goes in the opposite direction of most literature, which has focused rather on symmetry breaking, for obvious reasons. In the Standard Model (SM), electroweak and chiral symmetry are spontaneously broken as the universe cools \footnote{To be precise, in the SM both transitions are smooth cross-overs, made first order only by the addition of BSM ingredients.}. More generally, symmetry restoration at high temperature is a common theme in quantum field theory. 
Nonetheless, it is still possible for the universe to be stuck in a false vacuum of broken symmetry, as sketched in \cref{fig:sketch_symm_restoring}, and transition to a deeper, symmetric one. We discuss this further and study a simple, concrete two-field model in \cref{app:ToyModel}.

We start first by emphasising why the dynamics of such transitions are potentially interesting, or at least deserving of some consideration. We take symmetry restoring PTs as the prime example of more general transitions where particles lose mass as they cross the bubble wall. This is true by definition for gauge bosons (and typically fermions in the theory as well). If the bubble expands into the broken phase at highly relativistic speeds, we can think of it interacting with individual particles. In the frame of the wall, those particles are highly boosted and by simple kinematics each degree of freedom (d.o.f.) $i$ leads to a negative momentum transfer to the wall 
\begin{align}
\label{eq:DeltaP_LO}
    \Delta p_i = p_i^z - \tilde{p}_i^z \approx  -{  m_i^{2}\over 2 p_i^z }\ ,
\end{align}
where $p^\mu, m_i$  are the incoming particle's $4-$momentum and mass in the old (broken) phase, while a tilde will always denote a quantity in the new (symmetric) phase. We have in mind $ m_i = g v$ or $y v$ where $v$ is the symmetry breaking vev on the broken 
side, $g$ is the gauge coupling and $y$ is a Yukawa coupling. Multiplying by the incoming flux (assuming a thermal population at temperature $T$) gives the B\"odeker-Moore result \cite{Bodeker:2009qy} for relativistic pressure at leading order \footnote{We are ignoring numerical coefficients in front, made precise by \cref{eq:LO_pressure_full}.} 
\begin{align}
\label{eq:LO_pressure_Intro}
    {\cal P}_{LO} \sim -m_i^2 T^2 \ ,
\end{align}
indicating a preference towards continued accelerated expansion of bubbles.
This suggests that the phenomenology of these `inverse' PTs can be quite different and the strength of signals potentially enhanced. 
It is thus necessary to examine whether there are other effects, in the relativistic regime, that counter (or reinforce) this dynamic. 

\begin{figure}
\centering
    \includegraphics[width=.6\textwidth]{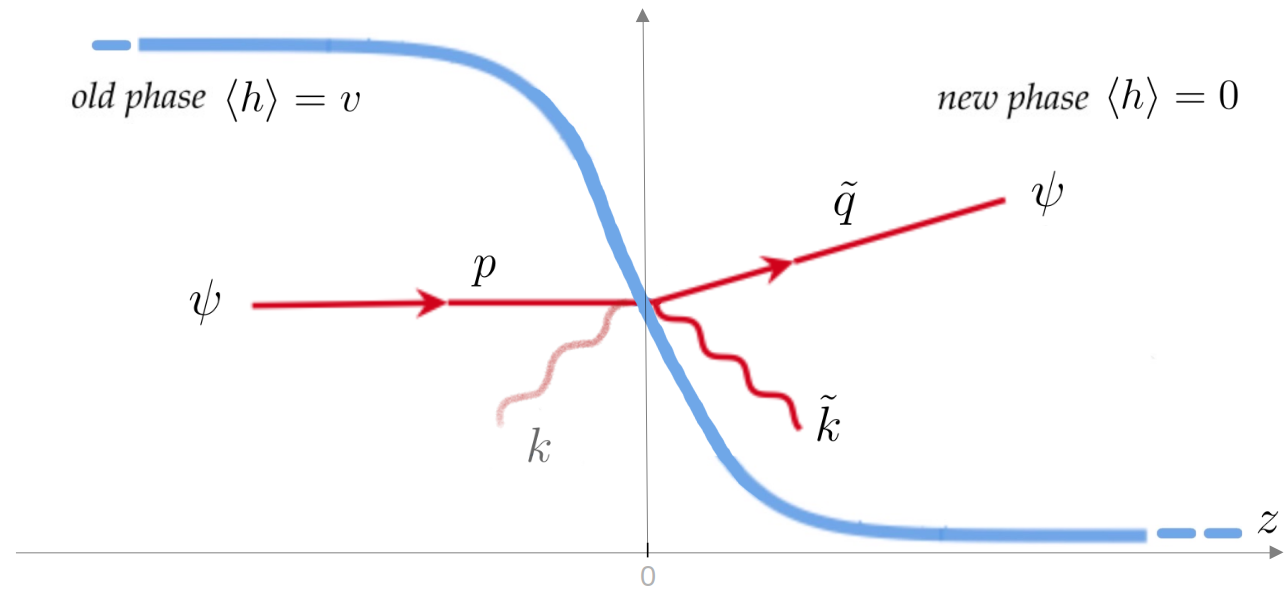}
    \caption{Our domain wall (in blue) describing the Higgs' vev $\vev{h}=v(z)$ is centered at $z=0$ in its rest frame, interpolating between a false vacuum phase of broken gauge symmetry and a true vacuum phase of restored symmetry. Highly relativistic medium particles hit the wall from the left and can undergo many $z-$momentum violating processes. In this work, we focus on a charged current crossing the wall while emitting a single quanta through the minimal coupling interaction in \cref{eq:Theory}. Asymptotically, the emitted particle can be a transverse gauge boson or a Higgs/Goldstone degree of freedom if emitting to the right, or any of the three polarisations of the corresponding massive vector if emitting to the left. Globally, the emitted degrees of freedom are best described by so-called $\tau$ and $\lambda$ polarisations, as presented in \cref{eq:Unitary_and_wall}. We use a tilde to denote quantities in the new phase.}
    \label{fig:MasterProcess}
\end{figure}

In this paper, we focus on friction pressure coming from the soft vector emission from a charged current, as it passes from a broken to a symmetric phase, as sketched later in \cref{fig:MasterProcess}. It is `NLO' in the sense that it is a $1\to 2$ process involving an interaction vertex (proportional to the gauge coupling $g$). For the more commonly studied symmetry breaking PTs, the analogous process (emission from a charged current going from symmetric to broken phase) leads to \cite{Bodeker:2017cim,Vanvlasselaer:2020niz,Gouttenoire:2021kjv,Azatov:2023xem}
\bea
{\cal P}_{NLO}\propto g^3 v \gamma_w T^3\ .   \qquad (\text{symmetric} \to \text{broken})
\eea
The growth with $\gamma_w$ means that in those PTs, this contribution eventually dominates over $1\to 1$ processes in the highly relativistic limit. Clearly, it is, therefore, crucial to understand these effects also in the symmetry restoring case. \\

For concreteness, we will be considering an Abelian Higgs model, with a complex scalar $H$ charged under a gauged $U(1)$ that is restored in the true vacuum. The Lagrangian is
\begin{align}
\begin{split}
    \label{eq:Theory}
    {\cal L} &= -\frac{1}{4} F_{\mu \nu} F^{\mu \nu}+| D_\mu H|^2 -V\left(\sqrt 2 |H|\right)+g J^\mu(\psi) A_\mu + \dots \ , \\
\text{and}~~~\vev{|H|}=0 ~~~&\hbox{ in true vacuum}\ ,~~~
\vev{|H|}\neq 0 ~~~\hbox{in false vacuum} \ ,~~~ \text{for some }T_{\rm nuc} \ ,
\end{split}
\end{align}
where $D_\mu \equiv (\d_\mu +ig A_\mu)$, $A^\mu$ is the gauge field, $V$ is a potential for $H$, and $J^\mu$ is a charged current made of fermionic or (complex) scalar matter $\psi$, minimally coupled \footnote{Finally, the dots in \cref{eq:Theory} will be completely irrelevant for us, but are there to allow for possible other d.o.f.s contributing to the realisation of an effective (in general finite-$T$) potential drawn in \cref{fig:sketch_symm_restoring}.}. 
We will study friction effects involving all single quantum emissions resulting from the coupling $\propto J^\mu A_\mu$. In the main text, we will eventually take the mass of this charged d.o.f. to be constant $m_\psi(z)=m_\psi$, ignoring any (Yukawa-like) interaction it might have with the Higgs field, and moreover will set it to zero (a good approximation in the relativistic limit of interest).
We comment on the absence of apparent soft divergences for this choice in \cref{app:IR div}, which is in contrast to the symmetry breaking case above. In  \cref{sec:Changing_current_mass} we argue that the effects of a spatially dependent $m_\psi(z)$  are subleading.\\

This work is organised as follows. In \cref{sec:TheorySetup} we present the setup of our calculation and review the canonical quantisation of the vector field in a background of broken translations \cite{Azatov:2023xem}, for completely general wall profile. In \cref{sec:Approximations_in_phase_space} we describe our strategy of using the step wall and WKB approximations in their respective, to some extent complementary, regimes of validity, without committing to a specific theory. The reader who is not interested in the many details of calculation can safely skip ahead to \cref{sec:results}, where we present our results, and more importantly to the final discussion in \cref{sec:summary}. In \cref{sec:Positivity_of_friction} we discuss more generally about the possibility of negative friction.

\section{Theoretical setup}
\label{sec:TheorySetup}

We imagine that a bubble of true vacuum $\vev{|H|}=0$ is nucleated and expands to large (compared to its initial) radius. Since we will be interested in the local physics of medium particles interacting with the bubble wall,
we will from now on consider the planar limit of a one dimensional domain wall $\vev{|H|} = v(z)/\sqrt{2}$ interpolating between the true vacuum at $z\rightarrow \infty$ and false vacuum $\vev{|H|} = v = \rm const $ at $z\rightarrow -\infty$. Moreover, since we are taking $v(z)$ to be time-independent and using the Minkowski spacetime metric, we are therefore working in an assumed inertial rest frame of the wall. This will be self-consistent when the bubble reaches equilibrium and travels at constant speed \footnote{More generally, it will be also if the timescale of the process on the wall we are interested in is short enough that an instantaneous rest frame can be assumed.}. The setup (as well as the process we wish to calculate) is summarised in \cref{fig:MasterProcess}. 

Expanding  $H=\left(v(z)+h + i \, h_2\right)/\sqrt{2}$ in \cref{eq:Theory} gives us a theory of perturbative quantum fields in the background of the domain wall.
To calculate particle processes in this spatially dependent background, we start by properly defining and canonically quantising the free theory. We will now state and explain the key features, while for more details we refer the reader to \cite{Azatov:2023xem}. The impatient or busy reader may skip directly to \cref{sec:quantisation} for a succinct summary.

\subsection{Unitary gauge and `wall' polarisations:} 
\label{eq:Unitary_and_wall}

Despite the fact that the theory approaches the symmetric phase asymptotically on one side, we can make use of unitary gauge \cite{Farrar:1994vp,Azatov:2023xem}. The equations of motion (e.o.m.) of the quadratic action are then
\begin{align}
    &\partial^2 h= - V''(v(z)) \, h \ ,
\\
\label{eq:A_EOMquadratic}
&\partial_\nu F^{\mu \nu}=   g^2 v^2(z) A^\mu \ , \quad \implies \partial_\mu \left(v(z) A^\mu\right)=0 \ .
\end{align}
The first will describe particle excitations of the Higgs $h$ with mass $m_{h,s}\equiv V''(0)$ and $m_{h,b}\equiv V''(v)$ on the symmetric and broken sides respectively \footnote{As well as two discrete modes describing fluctuations in position and shape of the wall, see for example \cite{Blanco-Pillado:2020smt}.}.
In this work we will only need to focus on the second - spatially dependent - Proca equation, which implies the generalised Lorentz condition in \cref{eq:A_EOMquadratic}. The excitations of $A^\mu$ will describe three massive vector d.o.f.s deep in the broken phase, while only two (transverse) polarisations deep in the symmetric phase, \textit{as well as the second Higgs d.o.f. $h_2$}. 
Since time and $\vec{x}_\perp=(x,y)$ translations remain symmetries of the theory, we can seek solutions to the quadratic e.o.m. that are eigenmodes of energy $k^0$ and transverse  (w.r.t. the wall) momentum $\vec{k}_\perp $
\begin{align}
    A^\mu_{k} = e^{-i (k_0 t - \vec{k}_\perp \vec{x}_\perp)} \;\zeta_{k}^\mu (z)  \ .
\end{align}
Using rotational symmetry around the $z-$axis we will always focus on 
\begin{align}
\label{eq:kperpChoice}
    \vec{k}_\perp = (k_\perp, \ 0) \ , \quad \text{with}  \ k_\perp>0  \ ,
\end{align}
unless otherwise stated.
The $z-$dependence of the background leads to a non-trivial part of the mode functions $\zeta^\mu(z)$, which are \textit{not} eigenmodes of $z-$momentum. Nonetheless one can choose boundary conditions such that $\zeta^\mu(z)$ describes the mode of choice asymptotically far away from the wall, with the full function everywhere else determined by solving the e.o.m..

A further complication in the presence of a spatially-dependent vector mass is that conventional transverse and longitudinal polarisations are in general not the best d.o.f.s to use. This is because the conventional spin angular momentum is associated with rotations around a particle's direction of propagation, for example, $\vec{k}=(\vec{k}_\perp,k^z)$ for a particle incoming from $z\rightarrow-\infty$. The presence of the domain wall background at $z=0$ means that only rotations around the $z-$axis remain a symmetry, which implies that spin is not conserved unless the particle is exactly incident on the wall $\vec{k}_\perp=0$. Physically this means, for example, that an incoming particle with conventional transverse polarisation has a finite amplitude to reflect a longitudinally polarised vector \footnote{Since the full theory \cref{eq:Theory} is invariant under 
 the full Lorentz group, the angular momentum lost is in reality of course absorbed by the wall.}.

 The appropriate, mutually orthogonal d.o.f.s in the presence of a wall were first pointed out in \cite{Farrar:1994vp}. We call them $\tau$ (of which there are two) and $\lambda$ `wall' polarisations. 
 The latter are easily obtained by setting $A_\tau^z=0$, which leads to $\partial_\mu A_\tau^\mu =0$ and $\left(\partial^2+g^2v(z)^2\right) A_\tau^\mu =0$, while the former have the orthogonal form $A_\lambda^\mu = \left(\partial^n \alpha(z), A^z\right)$, with $n=t,x,y$, where $\alpha(x)$ is fixed in terms of $A^z$ through the generalised Lorentz constraint in \cref{eq:A_EOMquadratic}.
 The $\zeta^\mu_{\ell = \tau_1,\tau_2,\lambda}(z)$ can be constructed from scalar functions $\tau_{i=1,2}(z)$ and $\lambda(z)$ as follows
\begin{align}
 \label{eq:Zeta Construction Tau} 
    \zeta_{\tau_i}^\mu &= \epsilon_{\tau_i}^\mu \; \tau_i(z) , \qquad \text{where} \quad \epsilon_{\tau_1}^\mu = (0, 0,1,0) \ , \qquad \epsilon_{\tau_2}^\mu = (k_\perp, k_0, 0, 0)/\sqrt{k_0^2 - k_\perp^2} \ , \\
        \zeta_{\lambda}^\mu &=\l( \frac{-i k^n \d_ z (v(z) \lambda)}{g E v^2(z)},\frac{E}{g v(z)}\lambda\r) \overset{\rm on \ shell}{=}  \bar{\d}^\mu \left({\partial_z ( v(z) \lambda )\over  E g \; v^2(z)}\right) +  {g v(z) \over E} \lambda \;\delta^\mu_z \ , 
        \label{eq:Zeta Construction Lambda}  \\
        & \text{with} \quad   E^2\equiv k_0^2-k_\perp^2 \ , \equiv \quad \bar{\partial^\mu} \equiv (-i k^n,  \d^z) \ , \nonumber
\end{align}
where the scalars satisfy corresponding Schrodinger-like wave equations that can be derived from \cref{eq:A_EOMquadratic}:
\begin{align}
    \label{eq:Schrodinger Tau}
&\left(E^2+\partial_z^2 - g^2 v^2(z) \right) \tau_{1,2} = 0  \ , \\
    \label{eq:Schrodinger Lambda}
&\left(E^2+\d_z^2 -  U_\lambda(z) \right)\lambda = 0 \ . 
\end{align}
The potential for $\tau$ modes is simply a changing vector mass. On the other hand, $\lambda$ has a non-trivial effective potential 
\begin{align}
\label{eq:Potential_lambda}
\begin{split}
   U_\lambda(z) = & \; g^2 v^2(z)  - v(z)\;\d_z \left( \d_z v(z) \over v^2(z) \right)  \ \rightarrow 
\begin{cases}
    g^2 v^2 \ , & \text{as} \ z \rightarrow - \infty \\
    m_{h,s}^2 \ , & \text{as} \ z\rightarrow  + \infty  
\end{cases} \ ,
\end{split}
\end{align}
which exhibits the expected asymptotic behaviour describing a propagating vector in the broken phase and a Higgs particle in the symmetric phase. The limits in the last step are true for any interpolating solution $v(z)$ satisfying $\partial_z^2 v(z) = V'\left(v(z)\right)$ for arbitrary potential with minima at $v$ and $0$ \cite{Azatov:2023xem}.

\paragraph{Orthonormal basis: }
Having identified the appropriate d.o.f.s and reduced the equations of motion to Schrodinger-like form, we must choose a convenient complete basis of solutions to expand in and eventually quantise. 
In this work, we will only be interested in processes with particles described by $A^\mu$ in the final state. We therefore choose boundary conditions corresponding to \textit{outgoing} particles moving towards the right ($R$) and left ($L$) respectively
\begin{align}
    \label{eq: R out general}
     \phi^{\rm out}_{\ell,R,k}(z) & \longrightarrow  
    \sqrt{k^z \over \tilde k^z}\begin{cases}
     t^\lambda_{R,k} e^{-i k^z z}
             \ , & \text{as} \ z \rightarrow - \infty\\
             e^{-i \tilde{k}^z z} +  r^\lambda_{R,k}   e^{i \tilde{k}^z z}  \ , & \text{as} \ z \rightarrow + \infty 
    \end{cases} \ ,\qquad &(\kt^z>0) \\
\label{eq: L out general}
  \phi^{\rm out}_{\ell,L,k}(z) & \longrightarrow
    \begin{cases}
            e^{i k^z z} +  r^\ell_{L,k}  e^{-i k^z z}  \ , & \text{as} \ z \rightarrow - \infty \\
            t^\ell_{L,k} \; e^{i \tilde{k}^z z}  \ , & \text{as} \ z \rightarrow  + \infty
    \end{cases}  \ , \qquad &(k^z>0) \\
    \text{with}& \tiny \quad  k^0 =  \sqrt{k_z^2+\kp^2+m_{\ell}^2}= \sqrt{\kt_z^2+\kp^2+\mt_{\ell}^2} 
\end{align} 
where $\phi_{\ell} \equiv (\tau_1(z),\tau_2(z),\lambda(z))$, the $r_k, t_k$ coefficients are fixed by solving the equations of motion, and the factor in front of \cref{eq: R out general} is for normalisation.  The general orthonormality condition is 
\begin{equation}
\label{eq:LRorthonormal}
    \int_{-\infty}^\infty dz \; \phi_{\ell,I,k} \; \phi_{\ell,J,q}^* = 2\pi \delta_{IJ}\delta(k^z-q^z) \ , \qquad I,J \in \{R,L\} \ .
\end{equation}
While the solutions \cref{eq:  R out general,eq: L out general} are not eigenstates of $z-$momentum, wavepackets formed by the superposition of purely $R$ or $L$ movers do describe localised single wavemodes \textit{at late time} with definite positive or negative central $z-$momentum respectively \footnote{At early times instead they describe \textit{two} separate wavemodes incoming towards the wall which interfere in just the right way to describe a well defined particle at late time.}. At finite $z$ the behaviour of the solutions will of course depend on the details of the shape of $v(z)$. Before moving on we will consider the simple, exactly solvable case of a step function.

\subsection{Step function}
\label{sec:stepFunction}
In this approximation we consider
\begin{align}
\label{eq:StepWallProfile}
    v(z) = v \left[1 -  \Theta(z)\right] \ ,
\end{align}
where $\Theta(z)$ is the Heaviside theta function. The wavemodes are exactly solvable and determined by the appropriate matching conditions at $z=0$. For the two $\tau$ polarisations, these are simply the continuity of the function and its first derivative, leading to 
\begin{align}
    \label{eq: chi R}
     \left(\tau^{out}_{i,R,k}\right)^* & =  
    \sqrt{k^z \over \tilde k^z}\begin{cases}
     \frac{\tilde{k}^z}{k^z }t^\tau_{k} \; e^{i k^z z}
             \ , & z<0 \\
             e^{i \tilde{k}^z z} -  r^\tau_{k}   e^{-i \tilde{k}^z z}  \ , & z>0 
    \end{cases}\ ,   \qquad &(\kt^z>0)  \\
\label{eq: chi L}
  \left(\tau^{out}_{i,L,k}\right)^* & =
   \begin{cases}
            e^{-i k^z z} +  r^\tau_{k}  e^{i k^z z}  \ , & z<0 \\
            t^\tau_{k} \; e^{-i \tilde{k}^z z}  \ , & z>0 
    \end{cases} \ , \qquad &(k^z>0) 
\end{align} 
where
\begin{align}
    r^\tau_{k} &= \frac{k^z - \tilde{k}^z}{k^z + \tilde{k}^z}\ ,~~~ t^\tau_{k} = \frac{2 k^z  }{k^z +  \tilde{k}^z}  \ , \quad
      k_0 = \sqrt{k_z^2+g^2 v^2+k_\perp^2} = \sqrt{\kt^2_z+k_\perp^2} \ . \nonumber
\end{align}
The orthonormality condition \cref{eq:LRorthonormal} can be checked exactly.

For $\lambda$, the step function limit is more tricky since on one side $\vt=v(z>0)=0$. However, we can obtain the right answer by starting from a broken to broken transition $\vt\neq0$, as first studied in \cite{GarciaGarcia:2022yqb}, and then sending $\vt\rightarrow 0$ subsequently. The matching conditions are continuity of $v(z) \lambda(z)$ and $\lambda'(z)/v(z)$ at $z=0$, leading to
\begin{align}
    \label{eq: lambda L}
    \left(\lambda^{\rm out }_{ R,k} \right)^*& =  
    \sqrt{k^z \over \tilde k^z}\begin{cases}
     \frac{\tilde{k^z}}{k^z}t^\lambda_{k} e^{-i k^z z}
             \ , & z<0 \\
             e^{-i \tilde{k}^z z} -  r^\lambda_{k}   e^{i \tilde{k}^z z}  \ , & z>0 
    \end{cases} \ ,  \qquad &(\kt^z>0) \\ 
\label{eq: lambda R}
  \left(\lambda^{\rm out}_{ L,k} \right)^*& =
    \begin{cases}
            e^{i k^z z} +  r^\lambda_{k}  e^{-i k^z z}  \ , & z<0 \\
            t^\lambda_{k} \; e^{i \tilde{k}^z z}  \ , & z>0 
    \end{cases} \ , \qquad &(k^z>0) 
\end{align} 
and
\begin{align}
    r^\lambda_{k} &= \frac{\tilde v^2 k^z - v^2 \tilde{k}^z}{\tilde v^2k^z + v^2\tilde{k}^z} \ , \quad  t^\lambda_{k} = \frac{2 k^z v \tilde v }{\tilde v^2k^z + v^2\tilde{k}^z} \ , \quad
      k_0 = \sqrt{k_z^2+m^2_\lambda+k_\perp^2} = \sqrt{\kt_z^2+\mt^2_\lambda+k_\perp^2} \ , \nonumber
\end{align}
where $m_\lambda = g v$, and $\mt_\lambda = g \vt$ for $\vt \neq 0 $ must be discontinuously changed to $\mt_\lambda = m_{h,s}$ in the case $\vt=0$. 
Note
\begin{align}
    r^\lambda_k \rightarrow -1 \ , \qquad \text{for} \quad \vt \rightarrow  0 \ .
\label{eq:rkGoesto1}
\end{align}
Thus, as long as the step wall is a good approximation, $\lambda$ modes are totally reflected: $R$ and $L$ movers live completely on opposite sides of the symmetry breaking wall. If we regularised \cref{eq:StepWallProfile} with something like $v(z)=v\left(1-\tanh(m_h z/2)\right)/2$, which comes from 
the standard double-well quartic potential $V(h)=\lambda_h h^2(h-v)^2/4$ studied in \cite{Farrar:1994vp}, then the total reflection is trivial. The width of the wall is controlled by the mass of the Higgs $L_w \approx m_{h}^{-1}$ and particles with $k^z \lesssim L^{-1}_w$, for which the step function is a good approximation, do not have enough energy to make it to the other side. However, we stress that even for wall profiles with $m_h \rightarrow 0$ the reflective behaviour remains. A better interpretation of \cref{eq:rkGoesto1} is that $\lambda$ describes the would-be Nambu-Goldstone boson, that interacts derivatively with the background wall, as seen for example by the derivative terms in the potential $U_\lambda(z)$ in \cref{eq:Potential_lambda}.


\subsection{Amplitudes and exchanged momentum}
\label{sec:quantisation}
We choose to expand the field $A^\mu$ into a complete orthonormal basis of solutions to the quadratic equations of motion that are eigenmodes of energy $k^0$, transverse momentum $\vec{k}_\perp$ and describe outgoing particles of definite $z-$momentum,
\begin{align}
 \begin{split}
       & A^\mu =  \sum_{I,\ell}\int \frac{d^3k}{(2\pi)^3\sqrt{2k_0}} \left( a^{\rm out}_{\ell,I,k}\; A^\mu_{\ell,I,k} + h.c. \right) \ ,  \\
       & \text{with} \quad  A^\mu_{\ell,I,k} = e^{-i (k_0 t - \vec{k}_\perp \vec{x})} \;\zeta_{\ell,I,k}^\mu (z) \ ,
    \label{eq:AfieldQuantisation}
 \end{split}
\end{align}
where  $\ell=\tau_1,\tau_2,\lambda$ sums over the three different orthogonal `wall' polarisations, and $I=R,L$ stand for right ($z\rightarrow + \infty $) and left ($z\rightarrow - \infty $) moving particles. 
In practice the $\zeta_{\ell,I,k}$ are constructed by solving the Shr{\" o}dinger equations \cref{eq:Schrodinger Tau,eq:Schrodinger Lambda} with boundary conditions of \cref{eq: L out general,eq: R out general} (the latter defining the `out' choice) and finally plugging the solutions into \cref{eq:Zeta Construction Tau,eq:Zeta Construction Lambda}.
Upon quantisation, the associated Fourier operators create out states
\begin{align}
    \left| k^{\rm out}_{\ell, I} \right\rangle &\equiv \sqrt{2 k_0} \; (a_{\ell, I,k}^{\rm out})^\dagger  \left| 0 \right\rangle  \ , \qquad I \in R,L \quad \& \quad \ell \in \tau_1, \tau_2, \lambda \ .
\end{align}
and satisfy the usual creation/annihilation operator algebra. 
Wick's theorem follows from this as usual and one may go ahead and compute amplitudes for arbitrary particle processes of interest in the background of the wall.

We will be focusing on the relatively simple single emission from the minimally coupled charged current $J^\mu$ in \cref{eq:Theory}.
The tree level amplitude for this process is given by the following  expression
\begin{align}
\label{eq:Amplitude_general_wall}
&\langle  k_{\ell ,I}^{\rm out} \; q  |\: \mathcal{S} \:| p \rangle = (2\pi)^3 \delta^{(3)}(p^n-k^n-q^n) i  \mathcal{M}_{\ell ,I}  \ , \nn
& i \mathcal{M}_{\ell ,I}\equiv  \int d z \left(\zeta_{\ell,I,k}^\mu(z)\right)^* (p+q)_\mu e^{i (p^z-q^z) z}  \ ,
\end{align}
where $S=\exp\left( i \int d^4 x \: g A_\mu J^\mu \right)$ is the $S-$matrix and Right and Left mover emission are to be treated as separate processes. We emphasise that for a conserved current $J^\mu$, which is the case we will focus on in this work, we can ignore the first term in \cref{eq:Zeta Construction Lambda} since it is a total derivative. 
The total average momentum transfer is then given by summing over polarisations, as well as right and left contributions,
\begin{align}
    \vev{\Delta p} = \sum_{\ell,I}  \vev{\Delta p^{\ell}_I} = & \, \sum_{\ell} \ \vev{\Delta p^{\ell}_R}+\vev{\Delta p^{\ell}_L} \ ,
\end{align}
and integrating appropriately over all final state phase space, giving the following master equations:
\begin{align}
    \label{eq:master-int-L}
\vev{\Delta p^{\ell}_L} = &
 {1 \over 64 \pi^2} \int_0^{k^z_{\rm max}} dk^z \int_0^{k_{\perp, \rm max}^2}\ dk_\perp^2 \cdot \dfrac{ |\mathcal{M}_{\ell, L}|^2}{k^0 p^z q^z} \Delta p_L \ , \\
  \label{eq:master-int-R}
 \vev{\Delta p^{\ell}_R} = &
 {1 \over 64 \pi^2} \int_{0}^{\kt^z_{\rm max}} d\kt^z \int_0^{k_{\perp, \rm max}^2}\ dk_\perp^2 \cdot \dfrac{ |\mathcal{\tilde M}_{\ell, R}|^2}{k^0 p^z q^z} \Delta p_R \ ,
\\
\text{and} & \quad \Delta p_L=p^z-q^z+k^z \ , ~~~\Delta p_R=p^z-q^z-\tilde k^z \ , \nonumber
\end{align}
where $\kt^z_{\rm max} \, , k_{\perp, \rm max}^2$ are defined below, and we naturally choose to integrate in $k^z$ for left emission and $\kt^z$ for right emission, since the lower kinematic limit is zero no matter the hierarchy between $m_\ell$ and $\mt_\ell$. The form of \cref{eq:master-int-L} follows straightforwardly from our choice to second quantise and normalise modes of $A^\mu$ in the $k^z$ variable (see \cref{eq:LRorthonormal}). 
The reader can find an explicit derivation in terms of wave-packets in \cite{Azatov:2023xem}. There, for the case of symmetry breaking, all particles were gaining mass in the new phase and we also integrated \cref{eq:master-int-R} in $k^z$, with a lower limit. On the other hand, for a symmetry restoring transition, $k^z$ takes on imaginary values for very soft $\tau$ emission to the right \footnote{For $\lambda$ emission, which of $k^z\ , \kt^z$ have an imaginary branch depends on the relative sizes of $m_\lambda = gv$ and $\mt_\lambda = m_{h,s}$.}, so that $\kt^z$ is necessarily the better variable. Changing quantisation variable will only change the normalisation of the wavemodes, which translates into
\bea
\mathcal{M} \ \rightarrow \ \mathcal{\tilde M} \equiv \sqrt{ \tilde k^z \over k^z} \mathcal{M}\ ,
\eea
and thus in the final state phase space integration
\bea
\label{eq:kz-im}
d k^z | \mathcal{M}^2| \ \to \ d \tilde k^z |  \mathcal{\tilde M}^2| \equiv
d\tilde k^z\l|\frac{\tilde k_z}{k_z}\r|| \mathcal{M}^2| \ .
\eea
The upper limits of integration are in general given by
\begin{align}
   \begin{split}
        & k^z_{\rm max} =  \sqrt{(p_0-\tilde{m}_\psi)^2-m_{\ell}^2} \ , \quad  
    \tilde{k}^z_{\rm max} = \sqrt{(k^z_{\rm max})^2 + m_{\ell}^2-\mt_{\ell}^2} \ , \\ 
    &k_{\perp, \rm max}^2= \frac{ (p_0^2 +E^2-\tilde{m}_\psi^2)^2}{{ 4p_0^2}}-E^2 \ ,
   \end{split}
\end{align}
where one chooses to write $E^2 \equiv k_0^2 - \kp^2$ in terms of  $k_z^2+m_\ell^2  $ or $\kt_z^2+\mt_\ell^2$ for left and right emission respectively.
For gauge symmetry restoration we have of course $\tilde m_\tau =0$.  In the main text we will also take the charged current to be massless $m_\psi=\tilde m_\psi = 0$ and comment on the insensitivity of our results thereon.

If one is powerful enough to compute the exact wavemodes for a given wall
profile,  the exact amplitude and average momentum transfer for transition radiation can be calculated using the expressions provided here. However, we will now explain our strategy to obtain approximate expressions, ignoring details of the wall shape.

\section{Approximations in phase space}
\label{sec:Approximations_in_phase_space}

The main difficulty in calculating amplitudes and then ultimately $\vev{\Delta p}$  in the previous section is the computation of the $\zeta^\mu(z)$ wavemodes for a given wall profile $v(z)$. We could in principle simply try to use a step function. Not only is this a good theoretical exercise, but it is also a legitimate approximation for soft vectors. However, one should not over-rely on it. Properties of the $\lambda$ field, in particular, can be qualitatively quite different in the exact step function limit, such as the enhanced reflection, as studied already in \cite{GarciaGarcia:2022yqb} and then \cite{Azatov:2023xem}.
In practice we will then follow the procedure described in \cite{Azatov:2023xem}, where phase space was split into two regions: IR, where the step wall approximation is legitimately valid and UV, where instead the wave function of the vector bosons can be treated in a WKB approximation. 

\subsection{Wavemodes}
We now examine exactly when the step wall and WKB approximations are legitimate. The conditions will end up being slightly more refined than the simple cut used in \cite{Azatov:2023xem}.
We recall one more time the asymptotic masses for both $\tau,\lambda$ polarisations, defining $m\equiv g v$ :
\bea
\label{eq:masses}
(\tau): \begin{cases}
   m_\tau(z=+\infty)= 0  \\
     m_\tau (z=-\infty)= m  \\
\end{cases} \ ,
\qquad
(\lambda): \begin{cases}
   m_\lambda(z=+\infty) = m_{h,s} \\
m_\lambda (z=-\infty)=m 
    \end{cases} \ .
\eea

\paragraph{Step function limit (IR):} The background wall can be well approximated by a step function for wavemodes which change on scales longer than those of the background. This should be expected for particles with momenta much less than the inverse width of the wall
\begin{align}
    \label{eq:step}
{\rm step}:\ {\rm Max}\l[|k^z| \, , |\tilde k^z|\r] \ll L_w^{-1}\ .  
\end{align}
The absolute value is necessary, since for example for $\tau$ polarisation there is a phase space region where 
\bea
&&k_z^2=E^2- m^2 <0 \ , \quad E^2\equiv k_0^2-k_\perp^2=\tilde k_z^2 >0 \ .
\eea
In other words, even if the mode is decaying for $z < 0$ as an $\exp(-|k_z z|)$, step wall approximation is valid only if $L_w^{-1} \ll |k_z|$.  This constraint, when implemented in the integration limits, will in general cut both the extrema of integration.


\paragraph{WKB limit (UV):}

We now analyse the range of validity of the WKB approximation. In general, we are looking at the solutions of the Shr{\" o}dinger-like equation: 
\begin{align}
    \left( \partial^2_z -V(z)\right) \psi = -E^2 \psi, \quad \begin{cases}
        V(z)=g^2 v^2(z) &\text{for }\tau \hbox{ polarisations (see \cref{eq:Schrodinger Tau})}\\
    V(z)=U_\lambda(z) &\text{for }\lambda \hbox{ polarisations (see \cref{eq:Schrodinger Lambda})}
    \end{cases} \ .
\end{align}
We look for a solution of  the form  $\psi = e^{i \varphi(z)}$, which leads to 
\begin{align}
    \left( i \varphi''(z) - \varphi'(z)^2 - V(z) \right) = -E^2 \ . 
\end{align}
Then assuming  $|\varphi''| \ll |\varphi'^{\,2}|, \ |E^2-V(z)|  \ ,$
we arrive at:
\begin{align}
    &\varphi'^{\,2}  = E^2 - m^2(z) \equiv k_z^2(s) 
    \implies \varphi(z) - \varphi(z_0) =  \int_{z_0}^z ds \ k^z(s) \ .
    \label{eq:fofzsolution}
\end{align}
The wave function solution becomes:
\begin{align}
    \psi = A \ e^{i \int_{z_0}^z ds \ k(s) }  \ + \mathcal{O}\left(|\varphi''|\over  |\varphi'^2| \right).
\end{align}
This approximation is valid if
\begin{align}
    |\varphi''| \ll |\varphi'^{\,2}|& \implies |k'_z(z)| \ll |k_z^2(z)| \implies {|\d_z V(z)| \over 2 |k_z(z)|}  \ll |k_z^2(z)| \ .
\end{align}
We approximate the derivative of the potential as $\d_z V(z)\sim \Delta m^2/L_w$, then
we arrive at the following constraint:
\begin{align}
\label{eq:wkb-simple}
     \quad {\rm Min} \l[  |k^z| \, , |\kt^z| \r] \gg \left({|\Delta m^2| \over 2 L_w}\right)^{1/3}.
\end{align}

\subsection{Amplitudes} \label{sec:amplitudes}

Having defined the two complementary approximations for the wavemodes $\zeta^\mu(z)$, one can attempt to evaluate the amplitudes \cref{eq:Amplitude_general_wall}, which will then be valid in the corresponding regimes. This is completely straightforward for the case of the step function and exact expressions are presented below in \cref{eq:ampl}. For WKB, the  matrix element for R emission is
\begin{align}
\begin{split}
     i \mathcal{M}_\ell^{\rm wkb} &= \int d z \: \mathcal{V}(z) e^{i \int^z_0 ds \:  \Delta p^z(s) }  \ , \\
   \text{with} \quad  &\mathcal{V}(z) \equiv  \epsilon_{\ell,k}^\mu(z)  \: (p+q)_\mu \ , \quad \Delta p^z(s)\equiv (p^z-q^z - k^z(s)) \ , 
\end{split}
\end{align}
where the constant $\epsilon^{\mu}_{\tau_{1,2},k}$ were given in \cref{eq:Zeta Construction Tau}, and we just take $\epsilon^\mu_{\lambda,k}(z) = \delta^\mu_z g v(z)/E$   using current conservation to ignore the total derivative in \cref{eq:Zeta Construction Lambda}. We can then reduce $\mathcal{M}_\ell^{\rm wkb}$ to the form first explored by \cite{Bodeker:2017cim}
\begin{align}
\label{eq:M_wkb_red}
    i \mathcal{M}_\ell^{\rm wkb \ red.} \equiv \int_{-\infty}^0 dz \: \mathcal{V}(-\infty) \: e^{i  (p^z-q^z - k^z )z}  + \int^{\infty}_0 dz \: \mathcal{V} (\infty) \: e^{i  (p^z-q^z - \kt^z)z  } \ , 
\end{align}
plus contributions which are dependent on the specific shape of the wall.
These additional contributions were shown to be important only if $ {\rm Min}_s \left[\Delta p^z(s) \right] L_w \gg 1$, as long as for $\lambda$ emission we subtract the total derivative term in \cref{eq:Zeta Construction Lambda} \footnote{If not, this piece will grow with energy and, ignoring the contribution inside the wall, one fails to see the cancellation that must necessarily occur.   
For more general processes, one should check case by case. 
}.
However, in this regime 
the full matrix element should vanish since we recover translation symmetry and the emission process is forbidden (see for details \cite{Azatov:2023xem}).
Ignoring the contribution inside the wall, it is not always guaranteed that this suppression is properly captured. Thus, in practice, we always complement \cref{eq:M_wkb_red} by explicitly cutting out the region $\Delta p_z L_w \gg 1$ in the phase space integration.  
Finally, we do not consider left emission in the WKB approximation because there is no allowed phase space simultaneously satisfying all appropriate cuts.

For consistency of notations with \cite{Azatov:2023xem} we report all amplitudes for the vector field quantised in terms of the $k^z$ variable:
\begin{align}
\label{eq:ampl}
\begin{split}
        \mathcal{M}^{\rm step}_{\tau,L}  &= - i g\epsilon_{\tau_2 }^\mu (p+q)_\mu \left( \frac{1}{\Delta p_r } + \frac{r^\tau_{ k}}{\Delta p} -  \frac{t^\tau_{ k}}{\Delta \tilde{p}_r} \right) \ ,
\\
   \mathcal{M}^{\rm step}_{\tau,R}  &= - i g \epsilon_{\tau_2 }^\mu (p+q)_\mu \sqrt{ k^z \over \tilde k^z} \left[ \frac{\tilde k^z}{k^z} \frac{t^\tau_k}{\Delta p} -   \frac{1}{\Delta \tilde{p}} + \frac{r^\tau_k}{\Delta \tilde p_r}\right] \ ,
\\
    \mathcal{M}^{\rm wkb\ red.}_{\tau}  &= - i g \epsilon_{\tau_2 }^\mu (p+q)_\mu \left( \frac{1}{\Delta p } - \frac{1}{\Delta \tilde{p}} \right) \ ,
\\
     \mathcal{M}^{\rm step}_{\lambda,L} &= - i  
     \frac{g^2 v}{E} (p^z+q^z) \left[ \frac{1}{\Delta p_r} - \frac{1}{\Delta p}\right],
\\
   \mathcal{M}^{\rm step}_{\lambda,R}  & = 0\ ,
\\
    \mathcal{M}^{\rm wkb\ red.}_{\lambda} &=- i  \frac{g^2v }{E} (p^z+q^z) { 1\over \Delta p } \ ,
\end{split}
\end{align}
where we have ignored terms proportional to delta functions since they ultimately don't contribute, and 
\begin{align}
\begin{split}
& \Delta p \equiv p^z-q^z-k^z \ , \qquad \qquad \:  \Delta p_r\equiv \Delta p_L \equiv p^z-q^z+k^z \ , \\
& \Delta \tilde{p} \equiv \Delta p_R \equiv p^z-q^z-\kt^z \ , \quad \Delta \tilde{p}_r \equiv p^z-q^z+\kt^z \ .
\end{split}
\label{eq:Deltap_notations}
\end{align}
By $\tau$ here we mean $\tau_2$. Amplitudes for the emission of $\tau_{1}$ are zero. Because of rotational symmetry around the $z-$axis we can focus on the choice \cref{eq:kperpChoice}, which manifestly leads to $\epsilon_{\tau_1}^\mu(p+q)_\mu = 0$.

\subsection{Numerical procedure}
\label{sec:numerical_procedure}

Strictly speaking, the conditions \cref{eq:step,eq:wkb-simple} tell us that the respective approximations are valid only when the values of momenta are 
much less/greater than the corresponding thresholds. In practice in our calculation, we make the following approximation:
\begin{align}
    \label{eq:presc-int-step}
&{\rm step}:\ {\rm Max}\l[|k^z| \, , |\tilde k^z|\r] < L_w^{-1}. \\
&{\rm WKB}: \l({\rm Min}  \l[ k^z, \kt^z \r] > \left({|\Delta m_\ell^2| \over 2 L_w}\right)^{1/3}\r)\cap \l( {\rm Max}\l[|k^z| \, , |\tilde k^z|\r] > L_w^{-1}\r).
\label{eq:presc-int-wkb}
\end{align}
For the WKB approximation, we have added the second condition in order to avoid double counting, since
for thin walls (e.g. $mL\ll 1$ for $\tau$) there can be an overlap between the phase space regions defined by \cref{eq:step,eq:wkb-simple}. 
Instead, for thick walls we comment that there can be a part of phase space not covered by our prescription, when
\bea
\l({\rm Min}\l[k^z, \tilde k^z\r] < \left({|\Delta m^2| \over 2 L_w}\right)^{1/3} \r)\cap \l({\rm Max}\l[k^z, \tilde k^z\r] > L_w^{-1}\ \r) \ .
\eea
However, this extra contribution will never significantly change our results. 

We gather now for convenience all of the separate contributions to the momentum transfer in our calculation scheme:
\begin{align}
\begin{split}
    \label{eq:all contributions}
\vev{\Delta p_L^{\rm step}}=& 
\dfrac{1}{64 \pi^2} \int dk^z d k_\perp^2\cdot \dfrac{|\mathcal{M}_L|^2}{p^zq^zk_0} \ (p^z-q^z+k^z)
\ \Theta ({\rm step}) \ ,\\
\vev{\Delta p_R^{\rm step}}=& 
\dfrac{1}{64 \pi^2} \int d\kt^z d k_\perp^2\cdot \l|\frac{\tilde k_z}{k_z}\r| \dfrac{|\mathcal{M}_R|^2}{p^zq^zk_0} \ (p^z-q^z-\tilde k^z)
\ \Theta ({\rm step}) \ ,\\
\vev{\Delta p^{\rm wkb}}=& 
\dfrac{1}{64 \pi^2} \int d\kt^z d k_\perp^2\cdot \l|\frac{\tilde k_z}{k_z}\r|  \dfrac{|\mathcal{M}^{\rm wkb \; red.}|^2}{p^zq^zk_0} \ (p^z-q^z-\tilde k^z)
\\
& \qquad \times\Theta\l({\rm WKB}\r)\Theta\left(L_w^{-1}-(p^z-q^z-\tilde k^z)\right) \ ,
\end{split}
\end{align}
where the $\Theta({\rm WKB})$ and $\Theta({\rm step})$ are imposing the conditions in \cref{eq:presc-int-step,eq:presc-int-wkb}. In the WKB contribution an additional $
\Theta\left(L_w^{-1}-(p^z-q^z-\tilde k^z)\right) $ is imposed to enforce that the momentum loss is always less than $L_w^{-1}$, as expected from Fourier transformation properties (see discussion below \cref{eq:M_wkb_red}, and
\cite{Azatov:2023xem} for more details). The integration ranges are given by
\begin{align}
\begin{split}
        &\tilde k^z \in \l[0 \ , p_0 - m_\psi \r] \quad \& \quad  k_{\perp}^2\in\l[0 \ ,    \frac{  ( p_0^2 +\tilde k_z^2-m_\psi^2)^2}{4p_0^2}-\tilde k_z^2    \r] \ , \\ 
&k^z \in\l[0 \ , \sqrt{(p_0-m_\psi)^2-m^2} \r] \quad \& \quad k_{\perp}^2\in\l[0 \ ,\frac{ (p_0^2 +k_z^2+m^2-m_\psi^2)^2}{{ 4p_0^2}}-k_z^2-m^2\r]\ .
\end{split}
\end{align}
where, as mentioned before, we have taken $m_\psi=\tilde m_\psi$ to be constant, and moreover will set it to zero since it does not affect our results in the relativistic regime.

\section{Results}
\label{sec:results}

\begin{figure}
    \centering
    \includegraphics[width=.48\textwidth]{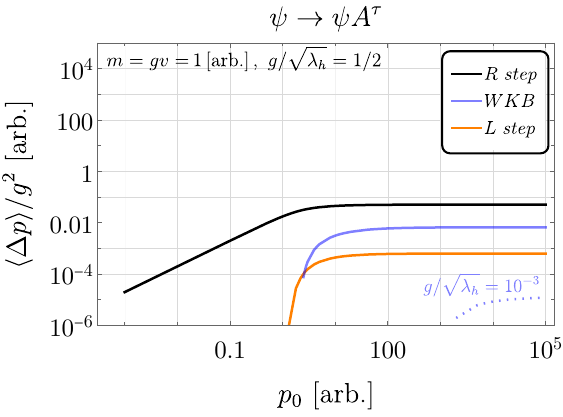}\includegraphics[width=.48\textwidth]{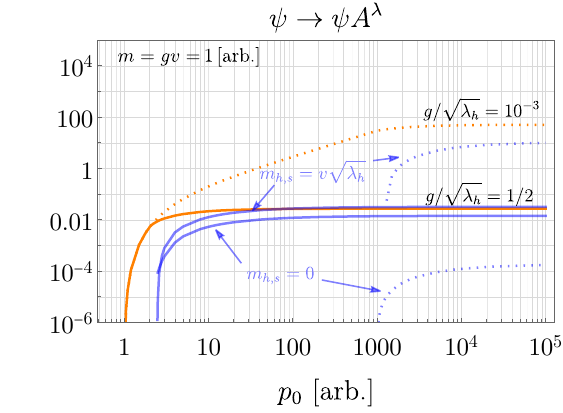}
    \caption{Average momentum transfer $\vev{\Delta p}$ from single emission by a charged particle with incoming energy $p_0=p^z$, as it crosses the wall into the phase of restored gauge symmetry. We show the breakdown of all contributions in our calculation scheme (see \cref{sec:Approximations_in_phase_space} for details): emission to the right ($R$) of low momentum modes (using the step function approximation), plus large momentum modes (using the WKB approximation), and emission to the left ($L$) of low momentum modes (step function). All are asymptotically constant, leading to friction on the wall that grows linearly with $\gamma_w$.
    We show results for two different values of wall length $m L_w = g /\sqrt{\lambda_h}= 1/2, 10^{-3}$, in units fixed by the broken phase gauge boson mass $m\equiv g v = 1$, described by solid and dotted lines respectively. \textbf{Left}: emission of $\tau-$polarisations. Only the WKB contribution changes appreciably in the thin wall limit, decaying as expected. \textbf{Right}: emission of $\lambda$ polarisations. $R$ emission in the step limit does not exist. $L$ emission depends linearly on $L^{-1}_w$. We also highlight the dependence of the WKB contribution on the mass of the Higgs in the symmetric phase $m_{h,s}$, showing the natural identification $m_{h,s}=L_w^{-1}=v\sqrt{\lambda_h}$ (upper two blue curves) as well as $m_{h,s}=0$ (lower two).}
    \label{fig:deltap vs p0}
\end{figure}
\begin{figure}
    \centering
   \includegraphics[width=.48\textwidth]{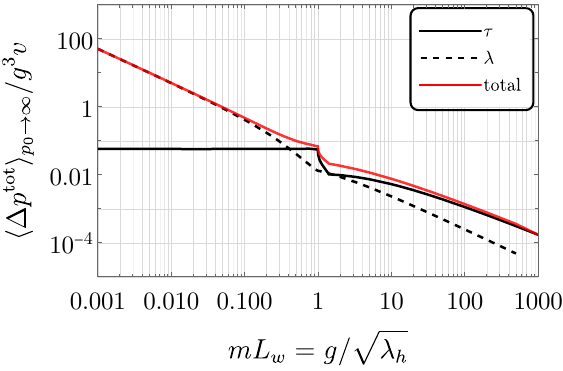} \includegraphics[width=.48\textwidth]{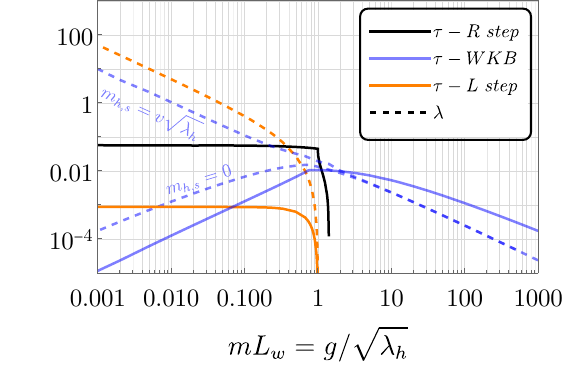}
    \caption{
    \textbf{Left}: Momentum transfer in the asymptotic limit, conveniently normalised, from $\tau$ and $\lambda$ emission channels, and the total sum, as a function of the product $mL_w$, expressed in terms of couplings as $g/\sqrt{\lambda_h}$. In the region $g/\sqrt{\lambda_h}\ll1$ (a `thin wall' from the point of view of the $\tau$ field), $\lambda$ is the dominant contribution. In the opposite (`thick wall') region we conservatively only keep WKB contributions. The jaggedness around $g/\sqrt{\lambda_h} \sim 1$ is purely a result of the brute cuts in phase space. \textbf{Right}: Breakdown of all contributions in our calculation scheme. Bold (dashed) lines refer to $\tau$ ($\lambda$) emission respectively. We show the $\lambda$ WKB contribution for the most natural identification $m_{h,s}=L_w^{-1}=v\sqrt{\lambda_h}$, as well as $m_{h,s}=0$.
    }
    \label{fig:deltap vs mL}
\end{figure}

In this section, we present our results for the average momentum transfer $\vev{\Delta p}$ from the NLO  process of single emission of $\tau$ and $\lambda$ polarisations from a charged particle in \cref{eq:Theory} traversing the wall. 

We break down all contributions as a function of incoming energy $p_0$ in \cref{fig:deltap vs p0}, where the main result is that $\vev{\Delta p}$ is constant and positive in the ultra-relativistic limit, and thus will lead to a pressure growing with $\gamma_w$ when multiplied by the incoming flux. In \cref{fig:deltap vs mL}, we explore in more detail the dependence of the constant asymptotic values $\vev{\Delta p}_{p_0\rightarrow \infty}$ on details of the theory. For $\tau$ emission the only scales in the problem (apart from $p_0$) are $m$ and $L_w$. Thus, by dimensional analysis, we have $\vev{\Delta p^{\tau}}_{p_0\rightarrow \infty} = g^3v f_\tau(mL_w)$, for some function $f_\tau$. We turn this into a dependence on couplings by defining $L_w^{-1} \equiv \sqrt{\lambda_h} \:v$, where $\lambda_h$ will be a coupling related to the Higgs potential. On the other hand, for $\lambda$ there is also a mass on the symmetric side $m_{h,s}$. In most models, it is natural to identify $m_{h,s} \approx L_w^{-1}$, which allows us to study $\lambda$ emission also as a function of $mL_w$. We thus take as a primary benchmark 
\begin{align}
    \label{eq:Identification_of_scales}
mL_w =\dfrac{g}{\sqrt{\lambda_h}}=\dfrac{m}{m_{h,s}} \ ,
\end{align}
but we comment on disentangling $L_w$ and $m_{h,s}$ when this make a significant difference.
We also recall that we take the emitting charged particle to have constant mass $m_\psi(z)=m_{\psi}$ and set it to zero. Turning on this mass distorts the curves at low energies but its effect is quickly lost for large $p_0$. In \cref{sec:Changing_current_mass} we argue that our estimates for total $\vev{\Delta p}_{p_0\rightarrow \infty}$ are also unaffected by allowing $m_\psi(z)$ to vary, although $\vev{\Delta p}$ can in general be even negative for low $p_0$.

We now provide some approximate formulae for all asymptotic contributions. Focusing first on $\tau$ emission, we obtain 
\begin{align}
    \langle \Delta p^{\rm \tau}_{R} \rangle_{p_0\to \infty} &\simeq {g^3v\over 8\pi^2}\begin{cases}
     4.4 \ , & g/\sqrt{\lambda_h} \ll 1\\
   0 \ , & g/\sqrt{\lambda_h} \gg 1
\end{cases}\ , \\ 
\langle \Delta p^{\rm \tau}_{L} \rangle_{p_0\to \infty} &\simeq {g^3v\over 8\pi^2}\begin{cases}
     0.07 \ ,& g/\sqrt{\lambda_h} \ll 1\\
    0 \ , &  g/\sqrt{\lambda_h} \gg 1
\end{cases} \ , \\
\langle \Delta p^{\rm \tau}_{wkb} \rangle_{p_0\to \infty} &\simeq {g^3v\over 8\pi^2}\begin{cases}
     g/\sqrt{\lambda_h} \ , & g/\sqrt{\lambda_h} \ll 1\\
    2  \left( g/\sqrt{\lambda_h}\right)^{-1} \ \ln \left(g/\sqrt{\lambda_h} \right) \ , & g/\sqrt{\lambda_h}  \gg 1
\end{cases}\ .
\end{align}
The first two contributions, coming from phase space where the step function is a good approximation, 
naturally dominate and remain constant in the thin wall limit $mL_w=g/\sqrt{\lambda_h}\ll 1$, while the contribution from the complimentary WKB regime vanishes. If instead one makes the wall look more and more broad $mL_w \gg 1$, in our scheme the WKB contribution dominates and is proportional to $g^2/L_w$, modulo an innocuous log. WKB domination follows from our cut in \cref{eq:presc-int-step}; if we got rid of the absolute value, the $R$ step contribution dominates over WKB.
However, we would like to emphasise that in the limit $\lambda_h\to 0\Rightarrow L_w\to \infty $ momentum 
transfer must vanish independently of prescription since the wall disappears, which is captured by the above equations. 
All contributions manifestly vanish as $g$ or $v \rightarrow 0$.
For $\lambda$ emission instead, we find
\begin{align}
\langle \Delta p^{\rm \lambda}_{L} \rangle_{p_0\to \infty} &\simeq {g^3v\over 8\pi^2}\begin{cases}
     4 \left(g/\sqrt{\lambda_h}\right)^{-1} \ , &  g/\sqrt{\lambda_h}\ll 1  \\
    0 \ , & g/\sqrt{\lambda_h} \gg 1
\end{cases}\ , \\
\langle \Delta p^{\rm \lambda}_{wkb} \rangle_{p_0\to \infty} &\simeq {g^3v\over 8\pi^2}\begin{cases}
     0.8\, \left(g/\sqrt{\lambda_h}\right)^{-1} \ , & g/\sqrt{\lambda_h}\ll 1\\
    1.6\,  \left(g/\sqrt{\lambda_h}\right)^{-1} \ ,  & g/\sqrt{\lambda_h}\gg 1
    \label{eq:DeltapLambdaWKB}
\end{cases}\ .
\end{align}
In the thin wall limit, the dominant contribution is from $L$ movers, and it grows linearly with $L_w^{-1}$. This can be traced back to the total reflection of $\lambda$ modes while the step function is a good approximation $k^z(z) \lesssim L_w^{-1}$, as discussed in \cref{sec:stepFunction}. 
For the natural identification $ m_{h,s}\approx L^{-1}_w$, then $\vev{\Delta p^\lambda}$ can be interpreted as going like the mass gained $m_{h,s}$. However, we stress that even if we set $m_{h,s}=0$, the behaviour remains.
A better interpretation is again that $\lambda$ describes the would-be Nambu-Goldstone boson, that interacts derivatively with the background wall. 
 The WKB contribution appears to also increase in the thin wall limit, perhaps contrary to the reader's expectations. This is only because of the identification $m_{h,s}\approx L^{-1}_w$ in our benchmark choice. If we set $m_{h,s}=0$, then we have instead
\begin{align}
     \langle \Delta p^{\rm \lambda}_{wkb} \rangle_{p_0\to \infty} \simeq {g^3v\over 8\pi^2} \, 2.2 \,\left(g/\sqrt{\lambda_h}\right) \ln\left((g/\sqrt{\lambda_h})^{-1}\right) \ , \quad \text{for} \ m_{h,s}=0 \quad  \text{and}\quad g/\sqrt{\lambda_h}\ll 1 \ .
\end{align}
where again we have $L_w^{-1} =\sqrt{\lambda_h}\: v$,
and it decays as expected as $\propto L_w^{-1}$. 
For the case $m_{h,s}\gg L_w^{-1}$ we found the WKB contribution to scale again like $L_w^{-1}$. This can be traced back to the fact that this limit is equivalent to the thick wall limit, happening for $\tau$ polarisation too.
For the benchmark \cref{eq:Identification_of_scales}, it is compensated by the mass gained $L^{-2}_w - m^2$,  and increases linearly instead, as per \cref{eq:DeltapLambdaWKB}. The difference is highlighted in both \cref{fig:deltap vs p0,fig:deltap vs mL}.
Finally, we recall that we do not have a contribution from R emission in the step function limit since the amplitude is proportional to $g \tilde v \rightarrow 0$, as already stated \cref{sec:amplitudes}.

In \cref{fig:deltap vs mL} the solid red line tracks the sum total of all contributions in our scheme, which we approximate by the following expression
\bea 
\langle \Delta p^{\rm tot} \rangle_{p^z\to \infty} &\simeq& {g^3v\over 8\pi^2}\begin{cases}
     5\, \left(g/\sqrt{\lambda_h}\right)^{-1} & g/\sqrt{\lambda_h}\ll 1\\
    2\left(g/\sqrt{\lambda_h}\right)^{-1}\l[ 1+ \ln\l( g/\sqrt{\lambda_h}\r)\r] & g/\sqrt{\lambda_h}\gg 1
\end{cases}\ ,
\eea
where we have switched to $p^z\rightarrow \infty$ to be more accurate.
To go to the friction pressure experienced by the wall expanding against a bath of particles charged under the restoring gauge symmetry, one should in principle integrate $\vev{\Delta p}$ against the distribution of incoming particles. Since this momentum transfer becomes constant for large $p_0$, to a very good approximation we can simply multiply \cref{eq:total friction} by the incoming flux.
Assuming a thermal distribution, this gives
\begin{align}
    \label{eq:total friction}
 {\cal P}_{\rm NLO}\simeq {\zeta(3) \over \pi^2} \gamma_w T^3 \langle \Delta p^{\rm tot} \rangle_{p^z\to \infty}\ , \quad \text{for} \quad \gamma_w \gg 1 \ .
\end{align}
This expression of course should only be trusted when the bubble can be thought of as interacting with individual particles. A necessary condition is thus that the timescale of interactions between particles (in the plasma frame) $\Gamma_{\rm int.}^{-1}$ is longer than the wall cross time, i.e. $\gamma_w \gg L_w\Gamma_{\rm int.}$.


\section{Summary and discussion}
\label{sec:summary}
In this paper, we studied for the first time NLO friction in phase transitions of gauge symmetry restoration, as a prime example of instances where particles may lose mass as they cross the phase boundary. To calculate, we reapplied the scheme developed in \cite{Azatov:2023xem} to study $1\rightarrow 2$ processes in the translation-breaking background of a bubble/domain wall, in particular interpolating between different phases of a gauge theory. This amounts to a first principles quantisation of the gauge field using appropriate polarisations, and splitting final state phase space into regimes where the step function and WKB approximations are respectively valid.

The main result of this paper is that the average momentum transfer due to a single emission from a particle charged under the gauge symmetry, as it passes from a broken to a restored phase, is positive, asymptotically constant, and roughly given by 
\begin{align}
\label{eq:SummaryDeltaP}
    \vev{\Delta p}_{p^z\rightarrow \infty} \sim {g^2 \over 2\pi^2}  L_w^{-1} \sim {g^2 \over 2\pi^2}  \sqrt{\lambda_h} \ v \ ,    \qquad \text{(symmetry restoring)}
\end{align}
which leads to a growing, positive, friction pressure \cref{eq:total friction} for ultra-relativistic bubbles. Thus, we prove that symmetry restoring phase transitions cannot have runaway behaviour $\gamma_w \rightarrow \infty$. 
    
\paragraph{Comparison with gauge symmetry breaking:} We can compare \cref{eq:SummaryDeltaP} to the far more studied case of symmetry breaking transitions \cite{Bodeker:2017cim,Vanvlasselaer:2020niz,Gouttenoire:2021kjv,Azatov:2023xem}, which is usually quoted as $\vev{\Delta p} \sim {g^3 v / 2\pi^2} $, ignoring a possible log term. 
A significant difference is present in the `thin wall' limit $mL_w = g/\sqrt{\lambda_h} \ll 1$ due to the enhancement of $\lambda$ ($\sim$ longitudinal) modes. 
The `thick wall' limit $g/\sqrt{\lambda_h} \gg 1$ instead has not been explicitly quoted in the literature, but it will also go as \cref{eq:SummaryDeltaP}, since an extremely broad wall $L_w\rightarrow \infty$ is like having no wall at all so that the average momentum transfer in this limit vanishes also for symmetry breaking transitions.
 Roughly speaking, one can imagine the 
black curve in the left panel of \cref{fig:deltap vs mL} as describing the symmetry breaking case. 
Another difference worth highlighting is the absence of naive soft divergences in any of the calculations in the main 
text. Unlike the symmetry breaking case, Nowhere did we have to resort to thermal masses to get a finite answer. We take a closer look at this in \cref{app:IR div}. When testing the insensitivity of our results to turning on a 
change in current mass $m_\psi(z)$ in \cref{sec:Changing_current_mass}, we did encounter a collinear log divergence, but this was always subleading with respect to the dominant contribution computed in the main text. 

\paragraph{Absence of negative friction:} Our motivation was guided partly by the possibility of negative friction in instances when particles lose mass as they cross the wall. In symmetry restoration, as 
studied in detail here, this is true for $\tau$ ($\sim$ transverse) polarisations of the gauge field \footnote{We note that, on the other hand, it is not in general true for $\lambda$ polarisations, whose mass changes from $m=gv$ to $m_{h,s}$.}. It is also true of any 
fermions with Yukawa couplings to the Higgs \footnote{Interestingly, a toy model with symmetry restoration discussed in \cref{app:ToyModel} leads to a positive \textit{total} ${\cal P}_{LO}$. It is therefore not always straightforward to obtain $\mathcal{P}<0$.}. However, we found by direct computation that contributions from $1\to 2$ NLO emission processes were positive and growing. This positivity is almost trivial. The only thing that can possibly be negative in \cref{eq:master-int-L,eq:master-int-R} are $\Delta p_{L,R}$, and it is easy to prove that both are $\geq 0$ for a constant charged particle mass $m_\psi = \tilde m_\psi$. Allowing $\psi$ to change mass in \cref{sec:Changing_current_mass} lead to a negative contribution that was however subleading. This was no accident.
In \cref{sec:Positivity_of_friction} we argue in generality that actually no particle physics process, with arbitrary number of incoming and outgoing states, can contribute a negative friction that is greater than that arising from the LO contribution $\vev{\Delta p} \approx p^z - p_0$ per d.o.f., in the limit $p^z \rightarrow \infty$. We do not rule out growing negative friction for transient regimes in special cases and leave the exploration of such a scenario, with its interesting though model-dependent phenomenological consequences, discussed briefly in \cref{sec:Positivity_of_friction}, to future study.
    

\section*{Acknowledgements}
 AA  and GB are in part  supported by the MUR contract 2017L5W2PT, AA also acknowledges the support from MUR project 20224JR28W.

\appendix

\section{Absence of soft divergences}
\label{app:IR div}
In this section, we elucidate why the total momentum transfer to the bubble wall in the case of a symmetry restoring phase transition calculated in the main text is IR finite.
This result should be contrasted with the opposite case (when the true vacuum has a broken gauge symmetry) where an apparent logarithmic IR enhancement has been found \cite{Azatov:2023xem, Gouttenoire:2021kjv}. Let us look at the expressions for the matrix elements see \cref{eq:ampl}, the divergences in principle can appear when the momentum differences in the denominator are vanishing 
\bea
\Delta \{p,p_r,\tilde p, \tilde p_r\}=0\ .
\eea
Let us consider for simplicity the case when the emitter particle has zero mass $m_\psi=0$ both in the false and true vacuum. Then the energy conservation forces
\bea
\Delta p_r&=&p^z-q^z+k^z> p_0-q_0-k_0=0\ , \\
\Delta  \tilde p_r&=& p^z-q^z+\tilde k^z \geq p_0-q_0-k_0=0\ ,
\eea
where $k_z,\tilde k_z$ are positive by definition and in the last inequality $\Delta p_r=0$ only for $k_0=0$. 
\bea 
\Delta p&=& p^z-q^z-k^z = p_0-\sqrt{q_0^2-k_\perp^2}-\sqrt{k_0^2-k_\perp^2 -m^2} > 0\ ,\\
\Delta \tilde p&=&p^z-q^z-\tilde k^z =
p_0-\sqrt{q_0^2-k_\perp^2}-\sqrt{k_0^2-k_\perp^2} \geq 0\ ,
\eea
where in the last inequality 
$\Delta \tilde p= 0$ only for $k_\perp=0$. So the divergences can appear only when $\Delta \tilde p$ and $\Delta \tilde p_r$ are equal to zero. Let us start by analyzing the divergences appearing in the limit $\Delta \tilde p \to 0$: the relevant amplitudes are $\mathcal{M}^{\rm step}_{\tau,R}$  and $\mathcal{M}^{\rm wkb}_{\tau}$ and both of them diverge as  $1/\Delta \tilde p$. 
For the scope of analyzing the IR divergences we can safely focus only on the divergent pieces. Below we show the calculation for the `step'  contribution (WKB contribution differs just by some finite factor and proceeds in exactly the same way). The amplitude squared in the $\Delta \tilde p \to 0$ limit is 
\bea
\lim_{\Delta \tilde p\to 0} |\mathcal{M}^{\rm step}_{\tau,R}|^2=\frac{4 k_\perp^2 p_0^2 |k_z|}{(\tilde k_z)^3}\times \frac{1}{(\Delta \tilde p)^2} \ .
\eea
However, what is controlling the momentum transfer in the wall is the quantity
\bea
\vev{\Delta p}|_{IR}\propto \int_{k_\perp\to0} \dfrac{d\tilde k_z}{2\pi} {1 \over 2k_0} \dfrac{dk_\perp^2}{4 \pi}\l|\frac{\tilde k_z}{k_z}\r|\cdot \dfrac{1}{2p^z
}\left[\dfrac{1}{2|q^z|} |\mathcal{M}_{\tau,R}^{\rm step}|^2 (\Delta \tilde p)\right]\ .
\eea
We change the integration variable using $\tilde k_z d \tilde k_z=k_0 dk_0$, and then focus only on the limit $k_\perp\to 0$,  where potential IR divergences can in principle appear. Then we get
\bea
\vev{\Delta p}|_{IR}\propto \frac{1}{4 \pi^2}\int \frac{d k_0}{k_0^2}k_\perp d k_\perp \ ,
\eea
where the integral over $k_\perp$ is obviously finite. The integral over $k_0$ looks naively IR divergent, but remember that the upper bound of the $k_\perp$ integration is always below $k_0$, thus
\bea
\vev{\Delta p}|_{IR}\propto \frac{1}{4 \pi^2}\int \frac{d k_0}{k_0^2}k_\perp d k_\perp < \frac{1}{8 \pi^2}\int d k_0= ~~~{\rm finite} \ .
\eea
Note that if we were calculating the probability of emitting the vector boson instead of the momentum transfer to the wall, we would have found a log divergence in $k_\perp$. Indeed,
\bea
\mathbb{P}_{1-\text{emission}}&\sim & \int_{k_\perp\to0} \dfrac{dk^z}{2\pi} {1 \over 2k_0} \dfrac{dk_\perp^2}{4 \pi}\cdot \dfrac{1}{2p^z
}\left[\dfrac{1}{2|q^z|} |\mathcal{M}_{\tau,R}^{\rm step}|^2 \right]\nn
&\sim &\frac{1}{4 \pi^2}\int \frac{d k_0}{k_0^2}k_\perp d k_\perp \times \frac{2 k_0 q_0}{p_0 k_\perp^2}= ~~~\hbox{Log divergent}\ .
\eea
What about $\Delta \tilde p_r =0$? This term appears in $\mathcal{M}^{\rm step}_{\tau,R}$ and in $\mathcal{M}^{\rm step}_{\tau,L}$. However, the contribution of the left moving modes will be non-singular since $k_0\geq m >0$, and for the right moving modes we get
\bea
\lim_{k_0\to 0}|\mathcal{M}^{\rm step}_{\tau,R}|^2\Delta \tilde p\propto  \frac{ m  p_0^2}{k_0^2}\ .
\eea
Thus, we are back to the estimate
\bea
\vev{\Delta p}|_{IR}\propto \int \frac{d k_0}{k_0^2}k_\perp d k_\perp =~~\hbox{finite}\ .
\eea


\section{Effects of a changing current mass}
\label{sec:Changing_current_mass}

In the main text, we computed the average momentum transfer for a single emission from a charged current crossing into a gauge symmetry restoring phase, assuming the charged matter did not change mass, i.e. $m_{\psi}(z)=m_\psi$. 
We then focused on $m_{\psi}=0$, as our results did not depend on this parameter for large $p_0$ (large $\gamma$). These were positive and straightforwardly IR-finite. 
Of course, the more realistic case will have a varying mass, and our results are only phenomenologically useful if they are not appreciably affected by it. We will now argue that this is indeed the case.

If we suppose for concreteness that the charged current is a fermion, then the most natural way to gain mass is by the addition to the theory in \cref{eq:Theory} of a Yukawa-like interaction
\begin{align}
\label{eq:fermionYukawa}
\delta{\cal L}_f &= \bar{\psi} i\cancel{D} \psi - (y H \bar{\psi}_L \psi_R + h.c.) \  , 
\end{align}
where gauge invariance forces $\psi$ to be exactly massless on the symmetric side. If $\psi$ were a complex scalar, we have more choice at the renormalisable level, such as 
\begin{align}
\label{eq:ScalarYukawa}
\delta{\cal L}_s = {1 \over 2}|D_\mu\psi|^2 - \tilde{m}^2_\psi |\psi|^2 \quad -(y\psi^2 H^2+h.c.)  \quad \text{or} \quad 
        y|\psi|^2 |H|^2 \ .
\end{align}
The first interaction mimics the fermion case, with current non-conservation upon expanding around the vev. The second term shows that for a scalar one can also gain mass in a current preserving fashion. All these differences will not matter in what follows. For concreteness, we focus on the maximal change $\tilde{m}_\psi=0$, in line with \cref{eq:fermionYukawa}.

A proper treatment would now require quantising also $\psi$ modes into left and right movers, as we did for the vector in \cref{sec:TheorySetup}. While this is certainly doable, it appears to us unnecessary, and we can argue qualitatively. 
On general grounds we expect the average momentum transfer to go as 
    \begin{align}
        \lim_{p_0 \rightarrow \infty}  \vev{\Delta p} \rightarrow v f\left(g,y,\sqrt{\lambda_h}\right) + \mathcal{O} \left( v^2/p^0 \right) \ ,
    \end{align}
    where $f$ is some unknown function. In the main text, we computed the leading $\mathcal{O}\left(g^3\right)$ and $\mathcal{O}\left(g^2 \sqrt{\lambda_h}\right)$ terms of $f$. This had dominant support from phase space where the vector was soft $k_0 \sim g v$, or $\sim L_w^{-1}, m_{h,s}$ for $\lambda$, and the fermion hard $q_0 \sim p_0$, consistent with our choice of ignoring the mass $m_\psi(z) \ll p_0$. Including $m_\psi(z)$, the greatest change will result in the part of phase space where the fermion is soft (and the vector hard).  
    
  Intuition for this possible contribution can be developed without much work by using the reduced WKB approximation for the matrix element, as per \cref{sec:amplitudes}. It was already pointed out in \cite{Bodeker:2017cim}  that the amplitude for emission of a soft fermion is suppressed with respect to the emission of a soft vector \footnote{See Table I in \cite{Bodeker:2017cim}.}. We now go a bit further, estimating this WKB contribution, and show that indeed the contribution related to a change in $\psi$ mass is suppressed for large incoming energies.

The matrix element for $\tau$ polarisation $\psi \rightarrow \psi A^{\mu}_\tau$ is
\begin{align}
\label{eq:MtauPsichanging}
       \mathcal{M}^{\rm wkb\ red.}_{\tau} &=- i y\epsilon^\mu_{\tau} \l(\frac{p^\mu+q^\mu}{p^z-q^z-k^z}-\frac{\tilde p^\mu+\tilde q^\mu}{\tilde p^z-\tilde q^z-\tilde k^z }\r) \ , 
\end{align}
where, as in the main text, by $\tau$ we mean the non trivially zero $\tau_2$. Defining $x=k^0/p^0$, the various four-momenta are explicitly
\bea
&&p^\mu=(p_0,0,0,\sqrt{p_0^2-m_\psi^2}) \, ,~~\tilde p^\mu=(p_0,0,0,p_0)\, ,\nn
&&q^\mu=(p_0(1-x),-k_\perp,0,\sqrt{p_0^2(1-x)^2-k_\perp^2-m_\psi^2})\ ,~~\tilde q^\mu=(p_0(1-x),-k_\perp,0,\sqrt{p_0^2(1-x)^2-k_\perp^2})\ ,\nn
&&k^\mu=(p_0x,k_\perp,0,\sqrt{p_0^2x^2-k_\perp^2-m_A^2})\ ,~~\tilde k^\mu=(p_0x,k_\perp,0,\sqrt{p_0^2x^2-k_\perp^2})\ .
\eea
The behaviour of \cref{eq:MtauPsichanging} is dictated by the collinear singularity of the second term for $k_\perp \rightarrow 0$, when
\bea
\mathcal{M}_{\tau}^{\rm wkb\ red.} \rightarrow iy\frac{4 p_0(1-x)}{k_\perp} \ . 
\eea
We note that the momentum transfer in this region is actually negative 
\bea
\Delta p_R = p^z-\tilde q^z-\tilde k^z \overset{k_\perp\to 0}{=} p^z-p_0 + \frac{k_\perp^2 }{2q^0 k^0} + \dots\ .
\eea
We can estimate this `negative' contribution from the loss in $\psi$ mass in the large $p_0$ limit by  
\bea
\vev{\Delta p_R}_{\psi}\propto \int \frac{d x dk_\perp k_\perp}{x p_0^2}\frac{x }{x (1-x)}\times \frac{16 p_0^2 (1-x)^2}{k_\perp^2}\times \frac{-m_\psi^2}{2 p_0} \ .
\eea
We observe that this integral diverges logarithmically with respect to $k_\perp$, but not with respect to $x$, because the WKB contribution is only considered in the UV part of the phase space, unlike the approach taken in Appendix \ref{app:IR div}, where we considered the step wall contribution. Nevertheless, the overall contribution must scale as
 \bea
 \vev{\Delta p}_{neg.}\propto -\frac{m_\psi^2}{p_0}\times \log (\hbox{divergence})\ .
 \eea
 This divergence should be considered regulated by some mass scale. If we had allowed the mass of the emitter in the new phase to be non-zero, we would have found a dependence such as
 \bea
 \vev{\Delta p}_{neg.}\propto - {m_\psi^2- \tilde m_\psi^2 \over p_0}\times \log \l( p_0 \over \tilde m_\psi  \r)\ .
  \eea
However, for large values of the initial momentum, this contribution will be highly subleading, regardless of the logarithmic divergence. Therefore, we can conclude that the contribution to the pressure we have identified is the only one that scales with the Lorentz boost factor. It is important to note that this divergence is related to the zero mass of the fermion on the symmetric side, and we expect it to be mitigated by interactions with the plasma.

We noticed that this negative contribution, coming from a particle losing mass across the wall, was subleading in $p_z\sim p_0$. This is true more generally as discussed in \cref{sec:Positivity_of_friction} where we explore the more general circumstances under which the splitting can result in negative friction.

\color{black}
\section{Negative friction?} 
\label{sec:Positivity_of_friction}

In this appendix, we attempt to examine more broadly the question of whether $1\to 2$ processes, or indeed any multi-particle process in theories of broken translation, can source negative friction. We should emphasise that, in order to be phenomenologically interesting, a negative sign is not enough, but rather we would require $\vev{\Delta p} \sim  (p^z)^{n}$ with $n>-1$. A contribution $n=-1$ is already obtained at LO when particles lose mass, while $n<-1$ would be inconsequential.

We start first with a generic $1\to 2$ process, with incoming particle $a$ and outgoing particles $b$ and $c$  and assign them $4-$momenta $p^\mu,q^\mu,k^\mu$ respectively. There are four possible physical momentum transfers, corresponding to particles $b,c$ being emitted to the right ($R$) or left ($L$) \footnote{As explained in \cref{sec:TheorySetup}, in theories of broken translations $R$ and $L$ emission are more properly considered on separate footing.}. Defining $\Delta p_{I,J}$, with $I,J \in \{R,L\}$, as appropriate for the case $b$ is $I-$emitted and $c$ is $J-$emitted, these are
\begin{align}
\label{eq:DeltaIJ_definition}
\begin{split}
        \Delta p_{R,R} &\equiv p^z-\tilde{q}^z -\tilde{k}^z  = \sqrt{p_0^2 - m_a^2} - \sqrt{(p_0 - k_0)^2 - \kp^2- \tilde{m}_b^2} - \sqrt{k^2_0 - \kp^2 - \tilde{m}_c^2 } \ , \\
    \Delta p_{L,R} &\equiv p^z-\tilde{q}^z + k^z 
    = \sqrt{p_0^2 - m_a^2} - \sqrt{(p_0 - k_0)^2 - \kp^2- \tilde{m}_b^2} + \sqrt{k^2_0 - \kp^2 - m_c^2 } \ , \\
    \Delta p_{R,L} &\equiv p^z + q^z - \kt^z  \ , \\
    \Delta p_{L,L} &\equiv p^z +q^z +k^z \ ,
\end{split}
\end{align}
where, as always, we use a tilde to denote a quantity evaluated in the new phase when it changes across the wall, and $q^z,\tilde{q}^z,k^z,\tilde{k}^z \geq 0$. Further down, we study each contribution in more detail, in particular when it can be negative. However, it is easy to prove that, in any case, and over all phase space, the momentum transferred cannot be smaller than the absolute minimum
\begin{align}
   \Delta p \geq \Delta p_{\rm Min} = p^z - q^0 - k^0 = p^z - p^0 =-{m_a^2 \over 2p^z} + \mathcal{O}\left( {1\over p_z^2}\right) \ , 
\end{align}
the second equality by energy conservation, which is not by accident equal to the leading order minimum \cref{eq:DeltaP_LO}. We note that, although this \textit{absolute minimum} is negative, it is going to zero in the asymptotic limit $p^z\rightarrow \infty$. It seems very difficult that the amplitude of the emission process might compensate for this. We can be more precise. A lower bound on the average momentum transfer is 
\begin{align}
\label{eq:proof}
   \begin{split}
        \vev{\Delta p_{I,J}} &= \int d\Pi_{\rm BTPH}\ |\mathcal{M}_{I,J}|^2 \ \Delta p_{I,J} \\ 
    &\geq \Delta p_{\rm Min} \int d\Pi_{\rm BTPH}\ |\mathcal{M}_{I,J}|^2  \\
    &= \Delta p_{\rm Min} \ \mathbb{P}_{I,J}\ ,
   \end{split}
\end{align}
where BTPH, which stands for broken translation phase space, is the positive-definite appropriate integration over final state momenta (such as in \cref{eq:master-int-L,eq:master-int-R}), and $\mathcal{M}_{I,J}$ and $\mathbb{P}_{I,J}$ are the matrix element and \textit{total integrated probability} for the corresponding process. Since $\mathbb{P}$ is an integrated probability for a physical process it cannot grow arbitrarily to infinity with incoming particle energy $p_0\approx p^z$ (less it break unitarity), and we conclude that 
\begin{align}
    \vev{\Delta p} > - {m_a^2 \over 2 p^z} \ , \qquad \text{for} \quad p^z \rightarrow \infty \ ,
\end{align}
that is, asymptotic friction (obtained when multiplying by the flux $\propto \gamma_w \propto p_z$) will never dominate the LO contribution. We saw an explicit example of a negative contribution being subdominant in \cref{sec:Changing_current_mass}.

The argument above can be easily run again for an arbitrary $N_i \to N_f$ process. In this case, the absolute minimum momentum transfer is 
\begin{align}
    \Delta p^{N_i\to N_f}_{\rm Min} = p_1^z + \dots + p_{N_i}^z - p_1^0 - \dots - p_{N_i}^0 =  - \sum_{j=1}^{N_i} {m_{a,j}^2 \over 2 p^z_j} + \mathcal{O}\left( {1\over p_z^2}\right) \ ,
\end{align}
which is independent of $N_f$. In the same way as \cref{eq:proof}, we would conclude that the average momentum transfer, over all final state channels, cannot beat the leading order contribution of $N_i$ particles simply crossing the wall, in the asymptotic limit.

We emphasise that our general argument applies to the \textit{asymptotic} regime. We do not see a reason to rule out negative friction for intermediate, even relativistic, bubble speeds. Perhaps thanks to processes mediated by higher dimensional operators that make $\mathbb{P}$ grow with $p_0$, until the cut-off scale.
We leave further study of possible intermediate regimes of growing negative pressure to future work. We simply point out that this could be phenomenologically quite interesting. Negative friction, scaling with a positive power of $\gamma$ could lead to instabilities. Fluctuations in the speed of expansion along the bubble's boundary can be expected, just as fluctuations in its position should be expected from finite temperature effects, collisions with the medium itself, or even quantum mechanically in the case of vacuum tunneling \cite{Blum:2024hcs}.
    These fluctuations would grow in time and could lead to highly asymmetric bubbles at late time, which would greatly enhance the gravitational wave signal from single bubble expansion as well as collisions \footnote{Similar conclusions are emphasised in a recent work \cite{Hassan:2024nbl}, although there the bubble of new phase is already nucleated with significant asymmetry.}.

\paragraph{Further details of $1 \to 2$ processes.}
We examine a bit further the contributions in \cref{eq:DeltaIJ_definition}. Clearly, $\Delta p_{L,L}>0$ always. Furthermore, $\Delta p_{L,R}$ is clearly minimised when all of the kinetic energy available goes to $b$ (i.e.  $k^z=0$) 
\begin{align}
\label{eq:DeltapLR}
    \Delta p_{L,R} \geq p^z - q^z_{\rm max}  &= \sqrt{p_0^2 - m_a^2} - \sqrt{(p_0-m_c)^2-\tilde{m}_b^2} \\
    & \rightarrow m_c -{ m_a^2 - \tilde{m}_b^2 \over 2 p_0} + \mathcal{O}\left(1/p_0^2\right) \ ,
\end{align}
and similarly
\begin{align}
\label{eq:DeltapRL}
    \Delta p_{R,L} \geq p^z - k^z_{\rm max}  &= \sqrt{p_0^2 - m_a^2} - \sqrt{(p_0-m_b)^2-\tilde{m}_c^2} \\
    & \rightarrow m_b -{ m_a^2 - \tilde{m}_c^2 \over 2 p_0} + \mathcal{O}\left(1/p_0^2\right) \ .
\end{align}
We see that the lower bounds \cref{eq:DeltapLR,eq:DeltapRL} can be positive or negative, depending on the relative size of the three masses involved, but if $p_0$ is large enough, they can only be negative if $m_{c,b}=0$ and $m_a>\tilde m_{b,c}$. 

Finally, the most negative possible momentum transfer comes from 
\begin{align}
    \Delta p_{R,R}   \geq  \left.\Delta p_{R,R} \ \right|_{\kp=0}  &= \sqrt{p_0^2 - m_a^2} - \sqrt{\left(p_0 - \sqrt{\tilde{k}_z^2+\tilde{m}_c^2}\right)^2 - \tilde{m}_b^2} - \tilde{k}^z \ , \\
    & \geq \sqrt{p_0^2 - m_a^2} - \sqrt{p_0^2 - \tilde{m}_{b,c}^2} \ , \quad \text{if} \ \tilde{m}_{c,b}=0 \ ,
\end{align}
where we have highlighted the simple case of when either of the outgoing masses is zero. When both are non-zero, the precise location of the absolute minimum within the allowed range $\kt_z\in (0, \sqrt{(p_0-\tilde{m}_b)^2-\tilde{m}_c^2})$ is non-trivially located somewhere in between the extreme points, but it is always greater than $p^z-p^0$.
\section{Toy model with symmetry restoring PT}
\label{app:ToyModel}


\begin{figure}
    \centering
    \includegraphics[width=.33\textwidth]{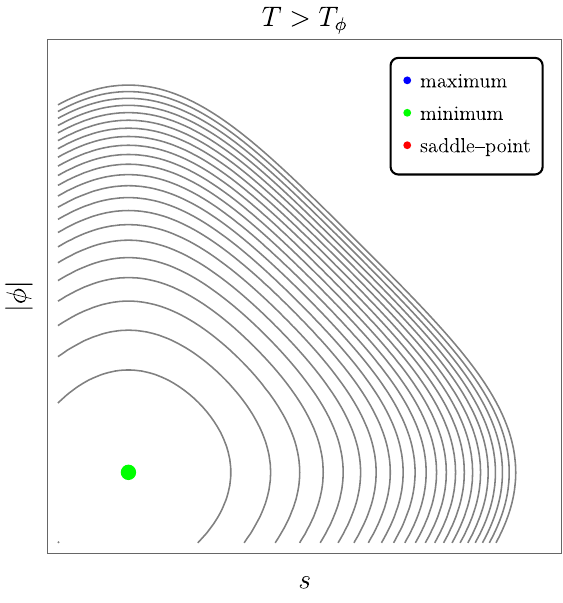}\includegraphics[width=.33\textwidth]{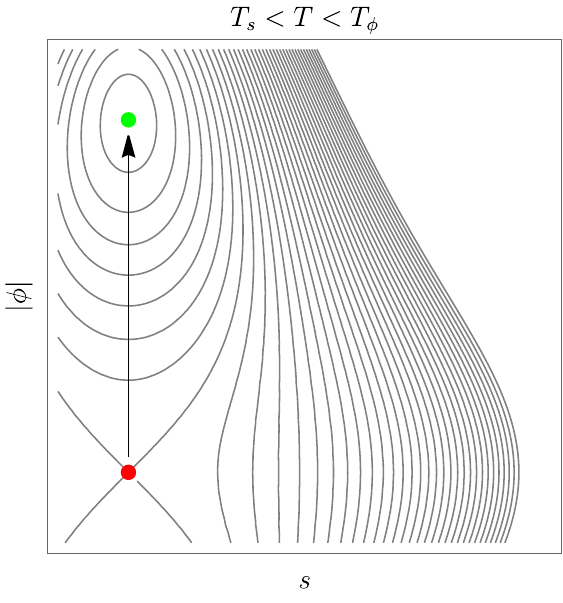}\includegraphics[width=.33\textwidth]{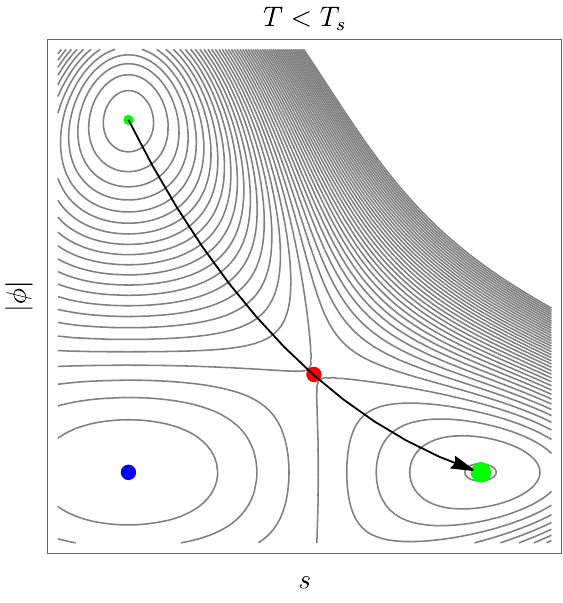}
    \caption{Sketch of the PT steps in the toy model described in \cref{app:ToyModel}, where $\phi$ is charged and $s\rightarrow -s$ is a symmetry. At early times, for very high temperature, there is only one, symmetry preserving, vacuum at the origin. As the universe starts to cool down, the origin becomes unstable and a new minimum appears, spontaneously breaking gauge symmetry. As the universe cools down further, a pair of two $\mathbb{Z}_2$ symmetry breaking minima appear (only one in the figure) and at some $T_{\rm nuc}$ the system will make the (1st order) transition to restore the gauge symmetry. The latter false and true vacua persist at zero temperature. Domain walls can in principle be avoided by introducing sufficient $\mathbb{Z}_2$ breaking. For the purposes of our work, we do not need to comment on this further.}
    \label{fig:sketch PT}
\end{figure}

In the main text, we computed friction without any discussion of specific models realising \cref{fig:sketch_symm_restoring}. 
We now dedicate some space to this, mentioning some general possibilities, before studying a simple concrete model below. 

In principle, for every symmetry breaking FOPT that proceeds as the universe cools down, there is a potential symmetry restoring FOPT that would follow if the universe were adiabatically heated up. This possibility was recently explored in \cite{Buen-Abad:2023hex}, in the context of post-inflationary reheating.
The picture in \cref{fig:sketch_symm_restoring} could also be true at zero temperature, for a non-renormalisable polynomial potential in (even powers of) the single field $|H|$ in \cref{eq:Theory}. Alternatively, by the addition of the right type of matter, one can arrange for radiative corrections to drive the effective quartic coupling of $h$ to run slightly negative and then positive (as occurs in the Standard Model - see for example \cite{Degrassi:2012ry}). This leads to the existence of a false vacuum at very large field value, which can be tuned to have higher energy density. 
For a two field (or more) potential, \cref{fig:sketch_symm_restoring} can be obtained even at the renormalisable level. In all these `cold' scenarios, one has to explain why the universe starts in the false vacuum, perhaps as a result of $h$ random walking during inflation. 

In the rest of this appendix, we explore a simple, though concrete, two field potential that can be considered as an existence proof for a first order, symmetry restoring phase transition that proceeds as the universe cools down, starting from a unique global minimum at high temperature. It does not therefore rely on inflation, or other initial condition selection mechanisms to start the universe in a false vacuum. For simplicity, we imagine the dynamics occurring in a hidden, thermal sector. For our purposes, we will not need to compute decay rates. We simply assume the transition will happen. We will however be interested in examining the leading order pressure mentioned around \cref{eq:LO_pressure_Intro}. We remind the reader how the LO pressure, in the high temperature limit, is defined for a general particle content
\begin{align}
\label{eq:LO_pressure_full}
    \mathcal{P}_{LO}= \sum_i {n_i c_i \over 24}\l(m_i^2(\text{TV})-m_i^2(\text{FV})\r) \, T_{\rm nuc}^2 \ ,
\end{align}
where $i$ runs over all the particles present in the thermal plasma, $n_i$ is the number of d.o.f.s of the particle $i$, $c_i=1(1/2)$ for bosons(fermions) while $m_i^2(\text{TV}, \text{FV})$ are the tree level field dependent masses computed at the true (false) vacuum. 
Interestingly, we will find it difficult for this particular scenario to lead to negative LO pressure.

\paragraph{Minimal concrete model:} We examine a simple, renormalisable, toy model consisting of a complex field $\phi$, charged under a gauged 
$U(1)$, and a real scalar $s$, endowed with a $\mathbb{Z}_2$ symmetry (a strong condition that we will later relax), with the following Lagrangian
\begin{align}
\label{eq:Toy_Model_concrete}
{\cal L}=- {1 \over 4} F_{\mu \nu}F^{\mu \nu}+|D_\mu\phi|^2+\frac{(\d_\mu s)^2}{2}+m_1^2|\phi|^2-\frac{\lambda_1 |\phi|^4}{4}+ \frac{m_2^2 s^2}{2}-\frac{\lambda_2 s^4}{4}-\lambda_3 s^2 |\phi|^2 \ .
\end{align}
It is obvious from the form of \cref{eq:Toy_Model_concrete} that, at zero temperature and for $\lambda_3$ large enough, one has a spontaneously broken $U(1)$ local minumum and a spontaneously broken $\mathbb{Z}_2$ local minimum:
\begin{align}
   \vev{|\phi|}=\sqrt{2}\: m_1/\lambda_1 \ , \vev{s}=0  \ ,  \qquad   
   \vev{|\phi|}=0 \ ,\vev{s}= m_2/\sqrt{\lambda_2} \ .
\end{align}
We will argue that, for some particular choices of parameters, the cosmology will follow the steps sketched in \cref{fig:sketch PT}. Initially, at very high temperature, the (only) minimum of the system is at the origin of the potential, that is the symmetric configuration $(\vev{|\phi|}=0,\vev{s}=0)$. Then, as the system cools down, $U(1)$ is spontaneously broken $ (\vev{|\phi|}\neq 0,\vev{s}=0)$. At even lower temperatures, a new, deeper minimum appears, which breaks the $\mathbb{Z}_2$ but restores the gauge symmetry, giving the desired transition $(\vev{\phi}\neq 0,\vev{s}=0)\to (\vev{\phi}= 0,\vev{s}\neq 0)$. 

For this to happen, we first need to explore the minima structure of the potential at low temperatures and require that the global minimum corresponds to $(\vev{\phi}= 0,\vev{s}\neq 0)$. This leads to the condition
\bea
{ m_2^4 \over 4\lambda_2} >\frac{m_1^4}{\lambda_1}\ .
\eea

For our discussion, it is sufficient to use the tree level potential, and the leading terms in high temperature expansion, wherein appear the field dependent masses. They read
\begin{align}
  \begin{split}
        m_{\phi_1}^2=&-m_1^2 +{\lambda_1 \over 4}(3 \phi_1^2+\phi_2^2)+\lambda_3 s^2 \ ,\\
m_{\phi_2}^2=&-m_1^2 +{\lambda_1 \over 4}( \phi_1^2+3\phi_2^2)+\lambda_3 s^2 \ ,\\
m_s^2=&-m_2^2+3\lambda_2 s^2+ 2 \lambda_3 |\phi|^2 \ , \\
m_A^2=&2g^2 |\phi|^2 \ ,
  \end{split}
\end{align}
where we separated two degrees of freedom of the complex scalar field $\phi=(\phi_1+i\phi_2)/\sqrt{2}$. Then, the thermal corrections to the masses will be given by
\bea
\Pi_{\phi}=\l(\frac{g^2 }{4}+\frac{\lambda_3+\lambda_1}{12}\r) T^2\ , \qquad \Pi_{s}=\l(\frac{\lambda_2}{4}+\frac{\lambda_3}{6}\r)T^2\ .
\eea
Therefore, the  requirement that $U(1)$ is broken before $\mathbb{Z}_2$ becomes 
\bea
T_\phi^2 > T_s^2\ ,
\eea
where $T_{\phi,s}$ are the temperatures when the effective mass, $m_{s,\phi}^2+\Pi_{s,\phi}$, vanishes and they are defined as
\bea 
T_\phi^2=\frac{12 m_1^2}{3 g^2+\lambda_1 +\lambda_3}\ ,\qquad
T_s^2=\frac{12 m_2^2}{3 \lambda_2+2\lambda_3}\ .
\eea 
This requirement, written in terms of the couplings, reads
\bea 
\label{eq: temp cond}
g^2<-\frac{\lambda_3+\lambda_1}{3}+\frac{m_1^2}{m_2^2}\l(\frac{2}{3}\lambda_3+\lambda_2\r)\ .
\eea
The last condition will be  that there is a barrier separating the  two minima at $(\vev{\phi}\neq 0,\vev{s}=0)$ and $ (\vev{\phi}= 0,\vev{s}\neq 0)$, that is we need to require that they are local minima, then
\bea
{\partial ^2 V_{\rm eff}\over \partial s^2}\biggr|_{\vev{\phi}\neq0,\vev{s}=0}=-m_2^2+\frac{4\lambda_3 m_1^2}{\lambda_1}+T^2\l[\frac{\lambda_2 }{4}-\frac{ \lambda_3(6 g^2+2 \lambda_3+\lambda_1)}{6 \lambda_1}\r]>0\ ,\\
{\partial ^2 V_{\rm eff}\over \partial \phi_1^2}\biggr|_{\vev{\phi}= 0,\vev{s}\neq0}=-m_1^2+\frac{\lambda_3 m_2^2}{\lambda_2}+T^2\l[ \frac{3 g^2-2 \lambda_3+\lambda_1}{12}-\frac{\lambda_3^2}{6\lambda_2}\r]>0 \ ,
\eea
where $V_{\rm eff}=V_{\rm tree}+V_T$. These conditions turn out to be easy to satisfy.

Let us look at the LO pressure in this model, focusing on its sign. Since all the particles here are bosons, all we need to calculate is the sum of the mass squared, then we get 
\bea
\sum m^2 (\vev{\phi}\neq 0,\vev{s}= 0)= \frac{12 m_1^2}{\lambda_1} g^2+ 2m_1^2 -m_2^2+\frac{4 \lambda_3 m_1^2}{\lambda_1}\ ,\\
\sum m^2 (\vev{\phi}= 0,\vev{s}\neq  0)= 2 m_2^2-2 m_1^2+ 2\lambda_3 \frac{m_2^2}{\lambda_2}\ ,
\eea
then the LO pressure will be negative if 
\bea
\sum m^2 (\vev{\phi}= 0,\vev{s}\neq  0)<\sum m^2 (\vev{\phi}\neq 0,\vev{s}= 0) \ ,
\eea
that translates into the condition
\bea
g^2 >- \frac{\lambda_3+\lambda_1}{3}
+\frac{\lambda_1m_2^2}{ 4\lambda_2 m_1^2}\l(\lambda_2+\frac{2}{3}\lambda_3\r)\ .
\eea
We can see that this is incompatible with the previous condition, in eq. \eqref{eq: temp cond}, of the $U(1)$ breaking before $\mathbb{Z}_2$ and the requirement that in the true vacuum, the gauge symmetry is unbroken. So the total LO pressure is always positive in the concrete theory of \cref{app:ToyModel}, despite the existence of a symmetry restoring phase transition. 

\paragraph{Including additional matter:} One might think that, in order to change the above conclusion, we need only introduce additional matter fields to the theory, that lose mass during the final symmetry restoring transition. Instead, we now show that, while maintaining our assumptions of renormalisability and $\mathbb{Z}_2$ symmetry, this is not so. We rewrite the theory in a slightly tidier fashion,
\begin{align}
\label{eq:Extended_toy_model}
    V=\lambda_f(f^2-v_f^2)^2+\lambda_t (t^2-v_t^2)^2+\lambda_{ft} f^2 t^2+ \sum (\hbox{other fields not getting vev}) \ ,
\end{align}
where $f$ can be thought of as the radial mode of the charged field and $t$ is the real scalar. The choice of labels reflects the fact that we will demand that, at $T=0$, the true (degenerate) vacua are $\vev t=\pm v_t$, $\vev f=0$, while the false vacuum is $\vev t=0$, $\vev f=v_f$. Again, both can be (meta-)stable local minima for large enough $\lambda_{ft}$. The desired hierarchy translates to the requirement
\bea
\label{eq:vacorder}
\lambda_f v_f^4 <\lambda_t v_t^4\ .
\eea
We take the additional sum in \cref{eq:Extended_toy_model} to be over an arbitrary number of extra fermions and scalars coupled at renormalisable level. For example, but not only, the Yukawa-like interactions in \cref{eq:fermionYukawa,eq:ScalarYukawa}. Again, we also restrict the sum to respect the $\mathbb{Z}_2$ symmetry of $t$, and relax this assumption further below.

As before, at the highest of temperatures, the origin of field space will be the unique, symmetric vacuum. 
Next, we impose that the $U(1)$ breaking occurs first, to follow the steps of \cref{fig:sketch PT}. Thermal corrections to the potential, in the high temperature limit, are given by \cite{Quiros:1999jp}
\bea
V_T(T)= T^2 \sum_i {n_i c_i \over 24} m_i^2(f,t) + \dots \ ,
\eea
where $m_i(f,t)$ are the field dependent tree level masses for every particle present in the thermal plasma, $n_i$ is the number of the d.o.f.s of the particle $i$, while $c_i=1(1/2)$ for bosons(fermions). Roughly speaking, the field $f$ will develop a vev at the temperature $T_f$,  defined as when its effective thermal mass at the origin vanishes
\bea 
m_{f}^{\rm eff}(f,t)^2\biggl|_{f,t=0}={\partial ^2 \over \partial f^2}(V_{\rm tree}+V_T(T_f))\biggr|_{f,t=0} =0 \ . 
\eea 
It is instructive to notice that the thermal correction to the mass can be rewritten in the following way
\bea 
{\partial ^2V_T \over \partial f^2}\biggr|_{f,t=0}=\Pi_f(T_f)={T_f^2} \sum_i {n_i c_i \over 24}\dfrac{\partial^2m_i^2(f, t)}{\partial f^2}\biggr|_{f,t=0} \ .
\eea
Now, whichever is the form of $m_i^2(f,t)$ after the second derivative, computed at the origin, the only terms that will survive are the quadratic ones, then it is convenient to write the masses as
\begin{align}
    \label{eq:expansion}
m_i^2(f,t)= m_i^2(0,0)+ {1 \over 2} {\partial^2 m_i^2(f,t) \over \partial f^2}\biggr|_{f,t=0}f^2+{1 \over 2} {\partial^2 m_i^2(f,t) \over \partial t^2}\biggr|_{f,t=0}t^2+\dots\ ,
\end{align}
where there is no linear term, or mixed second derivative, due to the $\mathbb{Z}_2$ symmetry. Restricting to renormalisable operators in \cref{eq:Extended_toy_model} means that there are no higher order terms and the series finishes here. We can then rewrite
\bea 
{\partial^2 m_i^2(f,t) \over \partial f^2}\biggr|_{f,t=0}=2\cdot{m_i^2(f,0)-m_i^2(0,0) \over f^2} \ .
\eea 
Therefore a useful way to write the thermal correction is the following
\bea 
{\partial ^2V_T \over \partial f^2}\biggr|_{f,t=0}=\Pi_f(T_f)={T_f^2} \sum_i {n_i c_i \over 12} \l(m_i^2(f, 0)- m_i^2(0,0) \over f^2\r) \ ,
\eea
Since the RHS is actually independent of $f$, this allows us to evaluate it wherever we want, so even at $f=v_f$. An analogous expression follows for $T_t$ evaluating at $t=\pm v_t$, and we can solve for both temperatures, finding
\begin{align}
\label{eq:TfandTt}
    T_f^2={-12 m_f^2(0,0)v_f^2\over\sum_i n_i c_i\l(m_i^2(v_f, 0)- m_i^2(0,0)\r)}\ , \qquad T_t^2={- 12 m_t^2(0,0)v_t^2\over\sum_i n_i c_i\l(m_i^2(0, \pm v_t)- m_i^2(0,0) \r)}\ ,
\end{align}
where we recall that $m_{f,t}^2(0,0)<0$. 
Now the condition $T_f> T_t$, together with eq.\eqref{eq:vacorder}, forces
\bea
\sum n_i c_i m_i^2(0,\pm v_t) - \sum n_i c_i m_i^2(v_f,0) > 0 \ ,
\eea
but this is the condition of the LO pressure being positive.

\paragraph{Explicit $\mathbb{Z}_2$ breaking:} Things become more complicated if we allow for explicit $\mathbb{Z}_2$ breaking terms in our theory. This could be simply done by including a cubic term in our tree level potential, or if one prefers to leave the cold potential unaltered, by adding to the sum of renormalisable operators in \cref{eq:Extended_toy_model}, some $\mathbb{Z}_2$ breaking terms, such as a massive Dirac fermion with a Yukawa coupling to $t$
\begin{align}
   \mathcal{L} \supset \ \mu \, t^3 + (y t  + m_\psi)\bar{\psi}\psi \ + \dots 
\end{align}
We find that with these types of additions it is naively possible to arrange for the cosmological history in \cref{fig:sketch PT}, with a negative LO pressure in the final symmetry restoring FOPT. However, this is at the expense of significant tuning, which will be very sensitive to a more proper treatment of finite temperature corrections to the potential, and thus cannot be trusted.

\color{black}
\bibliographystyle{JHEP}
{\footnotesize
\bibliography{biblio}}
\end{document}